\documentclass[11pt]{article}
\usepackage{epsfig}
\usepackage{rotating}
\usepackage{float,times}
\def\ref{\par\noindent\hangindent=6mm\hangafter=1}

\renewcommand{\theequation}{\arabic{enumi}.\arabic{equation}}

\begin{document}
\baselineskip 5.6mm

\begin{center}

{\bf  Large-scale Structures revealed by Wavelet Decomposition}\footnote
{The lecture of LZF at the 5th Current Topics of Astrofundamental
Physics, Erice, Sicily, 7-15 September 1996}

\bigskip
\bigskip
\bigskip
\bigskip

Li-Zhi Fang\footnote{Email address: fanglz@time.physics.arizona.edu}
 and Jes\'{u}s Pando\footnote{jpando@soliton.physics.arizona.edu}

\bigskip

Department of Physics\\
University of Arizona\\
Tucson, AZ 85721, USA
\end{center}

\newpage
\baselineskip 5.0mm

\tableofcontents

\newpage
\baselineskip 5.6mm

\section{ Introduction}

Large scale structure (LSS) study of Big Bang cosmology tries to explain how
an initially flat or smooth 3-dimensional surface described by the
Robertson-Walker metric evolved into a wrinkled one. In terms of density and
velocity fields, it explains how an initially homogeneous and
Hubble-expanding mass distribution evolved into its present inhomogeneous
state. It is generally believed that LSS was initiated by
fluctuations formed at the early universe, and that the subsequent clustering
was brought about by gravitational interaction between baryonic and dark matter
(Kolb \& Turner 1989). As a result, like the physics of dynamical critical
phenomena, turbulence, and multiparticle production in high energy collisions,
problems in LSS are typical of structure formation due to stochastic forces
and non-linear coupling (Berera \& Fang 1994, Barbero et al 1996).
The cosmic mass (or number) density distribution ${\rho(x)}$ can be
mathematically treated as a homogeneous random field.

Traditionally, the statistics, kinetics and dynamics in LSS are represented
by the Fourier expansion of the density field, $|\rho(k)|$. For instance, the
behavior of the LSS in scale space can effectively be described by the power
spectrum of perturbations $P(k) = |\rho(k)|^2$. In the case where the
homogeneous random field $\rho(x)$ is Gaussian, all statistical features of
$\rho(x)$ can be completely determined by the amplitude of the Fourier
coefficients. In other words, the two-point correlation function,  or its
Fourier counterpart the power spectrum, are enough to describe the
formation and evolution of the LSS.

However, the dynamics of LSS, such as clustering given by
gravitational instability, is non-linear. Even if the field $\rho(x)$
is initially Gaussian, the evolved density field will be highly non-Gaussian.
To describe the dynamics of the LSS knowledge of the phase of the
Fourier coefficients $\rho(k)$ is essential. As is well known, it is
difficult, even practically impossible, to find information about phase of
the Fourier coefficients as soon as there is some computational noise
(Farge 1992). This lack of information makes the description of
LSS incomplete. Even in the case where the phases are detectable,
the pictures in physical space, ${\rho(x)}$, and the Fourier space, $\rho(k)$
are separated. From the former we can only see the scales of the structures,
but not the positions of the considered structures, and {\it vice versa}
from the later. It has been felt for some time that the separate
descriptions between  Fourier (scale) and physical (position) spaces may
lead to missing key physics.

In order to resolve this problem, methods of space-scale-decomposition (SSD)
which might provide information about the phase (or position) and scale of the
considered structures have been developed. The possibility of
simultaneously localizing in both frequency (scale) and time (position) is
not new in physics. Anybody who listens to music knows that they, at any time,
can resolve the frequency spectrum. The problem of how to perform this time
resolution is also not new in physics. Wigner functions in quantum mechanics,
and the Gabor transform (Fourier transform on finite domain) were early
approaches.

Speaking simply, SSD represents a density field as a superposition of
density perturbations localized in both physical and scale spaces. For
instance, identification of clusters from a galaxy distribution by
{\em eyes} is a SSD. Generally, all methods of identifying clusters and
groups from surveys of galaxies or samples of N-body simulation are SSD.
One can list several popular SSDs in cosmology as follows: smoothing by a
window function, or filtering technique; percolation; the friend-to-friend
algorithm; count in cells (CIC).

A common problem of most of the above mentioned SSDs is that the
bases, or representations, given by these methods is incomplete. Unlike the
Fourier representation $\rho(k)$, these SSDs lose information contained in
the density field $\rho(r)$. For instance, one can completely reconstruct the
density field $\rho(x)$ by the Fourier coefficients $\rho(k)$, but cannot do
the same using window filters, CIC, percolation etc.

All these SSDs are, directly or indirectly, the precursors to the DWT
(Discrete Wavelet Transform). The DWT is also a SSD, but is based on
bases sets which are orthogonal and complete. The
DWT is invertible and admissible making possible
a complete representation of  LSS without losing
information. Unlike the Fourier bases (the trigonometric functions)
which are inherently nonlocal, the DWT bases have limited spatial
support. The DWT allows for an orthogonal and complete projection on modes
localized in both physical and space spaces and makes possible a
multiscale resolution.

Moreover, the orthogonal bases of the DWT are obtained by (space)
translation and (scale) dilation of one scale function (Meyer 1992, 1993;
Daubechies 1992). They are self-similar. This translation-dilation
procedure allows for an optimal compromise: the wavelet transform gives very
good spatial resolution on small scales, and very good scale resolution on
large scales. Therefore, the DWT is able to resolve an arbitrary density
field simultaneously in terms of its position variable and its conjugate
counterpart in Fourier space (wavenumber or scale) up to the limit of
uncertainty principle.

There have been attempts to use the continuous wavelet transform (CWT) to
analyze LSS (Slezak, Bijaoui \& Mars 1990; Escalera \& Mazure 1992; Escalera,
Slezak \& Mazure
1992; Martinez, Paredes \& Saar 1993). However, since 1992 it has become
clear that the CWT is a {\it poor} or even {\it impossible} method to use as  a
reasonable SSD. The difference between CWT and DWT is mathematically
essential, unlike the case for the Fourier transform, for which the
continuous-discrete difference is only technical (Yamada \& Ohkitani 1991;
Farge 1992; Greiner, Lipa \& Carruthers 1995).

These properties of the DWT make it unique among the various SSD methods.
One can expect that some statistical and dynamical features of LSS can
easily, and in fact {\it only}, be described by the DWT representation.
The DWT study of LSS now is in a very preliminary stage. Nevertheless,
results have shown that the DWT can reveal aspects of LSS behavior
which have not been seen by traditional methods (Pando \& Fang 1995, 1996a,
1996b; Huang et al 1996). These DWT-represented features have also been
found to be effective for discriminating among models of LSS formation.

As we will show the DWT opens a new dimension in the study of the
statistics and dynamics of the LSS.

\setcounter{enumi}{2}
\setcounter{equation}{0}

\section{Discrete wavelet transform of density fields}

The real difference in using the discrete wavelet transform (DWT) as compared
with say, Fourier techniques, comes when one deals with samples of finite
extent.  Since both the bases are complete, the information revealed by both
these techniques is equivalent when the function is continuous and the limits
of the respective inner products or sums can be calculated at infinity.
However, once the function is not continuous (or, rather, not continuously
sampled) or the sum cannot be calculated to infinity, the two representations
reveal different aspects of the distribution.  But this is always the case
when dealing with either observational data or simulated data.  Limited
resolution always forces one to sample a function at intervals, and no sum
can be calculated to infinity.  Unlike the Fourier transform, the
difference between the continuous and discrete transform is not merely
technical (Yamada \& Ohkitani 1991; Farge 1992; Greiner, Lipa \&
Carruthers 1995).  One does not merely replace the limit of a sum with
infinity and then take a limit.  By construction, the discrete and continuous
wavelet transform are quite different.

\subsection{An example: Haar wavelet}

Let us first consider a 1-dimensional (1-D) density field $\rho(x)$ over
a range $0 \leq x \leq L$. It is not difficult to extend the
1-D analysis to higher dimensions. It is convenient to use the density
contrast defined by\footnote{We do not as usual denote the density
contrast by $\delta$, because of possible confusion with the Kronecker
$\delta$ symbol}
\begin{equation}
\epsilon(x) = \frac{\rho(x) - \bar\rho}{\bar\rho}
\end{equation}
where $\bar\rho$ is the mean density in this field.
Actually, observed data and simulated samples can only provide density
distributions with finite resolution, say $\Delta x$. Hence, without loss
of information,
$\epsilon(x)$ can be expressed as a histogram with $2^{J}$ bins (Figure 1),
where $J$ is taken large enough so that $L/2^J < \Delta x$ i.e.
\begin{equation}
J \leq \bmod (|\ln\Delta x|/\ln2) +1,
\end{equation}

The histogram is labeled so that the $2^J$ bins are designated by an
integer, $l$, running from
$0 \leq l \leq 2^J-1$. Bin $l$ covers a range from $Ll2^{-J}$
to $L(l+1)2^{-J}$. The samples can fully be described by the $2^J$
$\epsilon_{J,l}$ ($0 \leq l \leq 2^J-1$) defined by
\begin{equation}
\epsilon_{J,l} = \epsilon(x),
\hspace{2cm} Ll2^{-J} \leq x \leq L(l+1)2^{-J}.
\end{equation}
Using $\epsilon_{J,l}$, we can rewrite $\epsilon(x)$ as
\begin{equation}
\epsilon(x) = \epsilon^{J}(x) \equiv
\sum_{l=0}^{2^{J}-1}\epsilon_{J,l}\phi_{J,l}^{H}(x)
\end{equation}
where the $\phi_{J,l}^{H}(x)$ are given by
\begin{equation}
\phi_{J,l}^{H}(x) = \left\{ \begin{array}{ll}
1 & \mbox{for $Ll 2^{-J} \leq x \leq L(l + 1) 2^{-J}$}\\
0 & \mbox{otherwise.}
\end{array} \right.
\end{equation}
Actually, $\phi_{Jl}^{H}(x)$ is a top-hat window function on
resolution scale
$L/2^J$ and at position $Ll 2^{-J} \leq x \leq L(l + 1) 2^{-J}$.

Expression (2.5) can be generalized to top-hat window functions on
different scales. We first define a top-hat scaling function as
\begin{equation}
\phi^{H}(\eta) = \left\{ \begin{array}{ll}
1 & \mbox{for 0 $\leq \eta \leq$ 1}\\
0 & \mbox{otherwise.}
\end{array} \right.
\end{equation}
Thus, one can construct a set of top-hat window functions by a
translation and dilation of the scaling function (2.6) as
\footnote{Actually, eq.(2.7) is not a dilation of eq.(2.6), but
a compression. We use the word ''dilation", because in the wavelet
literature the factor $1/2^{-j}$ is called the scale
dilation parameter, regardless if it is larger than 1.}
\begin{equation}
\phi_{j,l}^{H}(x)= \phi^{H}(2^{j} x/L - l),
\end{equation}
where $j$, $l$ are integers, and $j \geq 0$, $0 \leq l \leq 2^j - 1$.
Obviously,
$\phi_{j,l}^{H}(x)$ is a normalized top-hat window function on scale
$L/2^j$ and at the position $Ll 2^{-j} \leq x \leq L(l + 1) 2^{-j}$.
$\phi_{j,l}^{H}(x)$ is called the mother function.

If we smooth the density contrast $\epsilon(x)$ by the mother function
$\phi_{j,l}^{H}(x)$ on scale $j = J-1$, we have an approximate expression
of $\epsilon(x)$ as
\begin{equation}
\epsilon^{J-1}(x) = \sum_{l=0}^{2^{J-1}-1}\epsilon_{J-1,l}\phi_{J-1,l}^{H}(x)
\end{equation}
where the mother function coefficients (MFCs) $\epsilon_{J-1,l}$ are given
by
\begin{equation}
\epsilon_{J-1,l} = \frac {1}{2} (\epsilon_{J,2l} + \epsilon_{J,2l+1}).
\end{equation}
Similarly, we can continue this procedure to find smoothed distributions
$\epsilon^{j}(x)$ on scales $j=J-2, J-3,...$ as
\begin{equation}
\epsilon^{j}(x) = \sum_{l=0}^{2^{j-1}-1}\epsilon_{j,l}\phi_{j,l}^{H}(x)
\end{equation}
where the $j$-th MFCs can be found from $(j+1)$-th MFCs by
\begin{equation}
\epsilon_{j,l} = \frac {1}{2} (\epsilon_{j+1,2l} + \epsilon_{j+1,2l+1}).
\end{equation}

These results are nothing new. In fact they are window smoothing
on a scale-by-scale basis
as shown in Figure 1. The $\epsilon_{j,l}$ contain less information
than $\epsilon_{j+1,l}$. In order not to lose any information as a result of
the smoothing, we should calculate the difference between the smoothed
distributions on succeeding scales, say $\epsilon^{j+1}(x)- \epsilon^{j}(x)$.
Figure 1 also plots these differences.
\begin{figure}[htb]
\centering
\epsfig{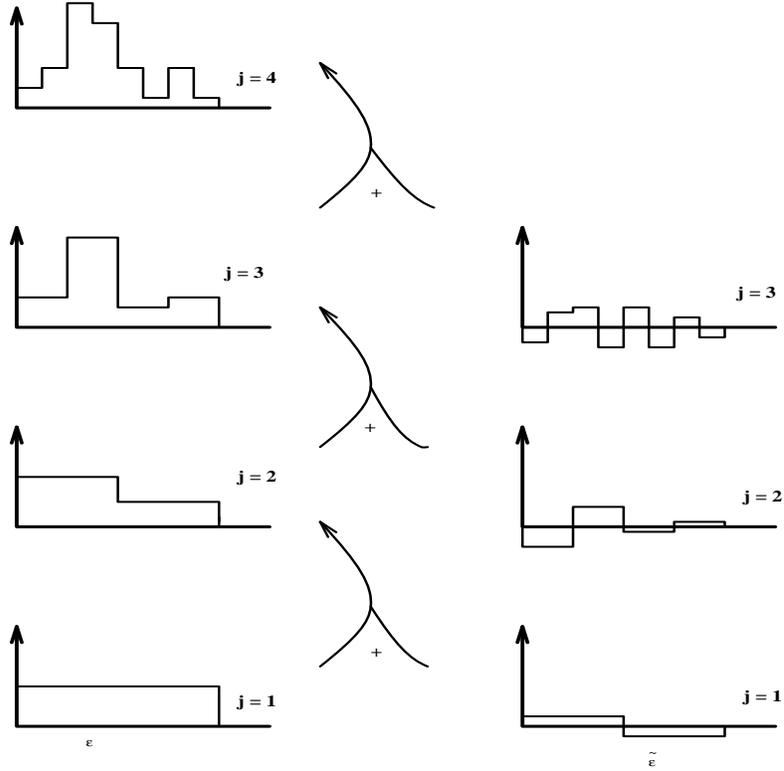}
   \label{fig1}
\caption{A Haar wavelet multiresolution decomposition.
The original sample is shown by the top left figure. Its resolution
is $j =$ 4. The left column is the reconstructed distributions of the
MFCs $\epsilon_{j,l}$ on scales $j=$ 3, 2, 1. The right column of
histograms show the distribution of FFCs, $\epsilon_{j,l}$ on
the corresponding scale $j$.}
\end{figure}

To describe this difference, we define a wavelet as
\begin{equation}
\psi^{H}(\eta) = \left\{ \begin{array}{ll}
1 & \mbox{for $0 \leq \eta \leq 1/2$} \\
-1 & \mbox{for $1/2 \leq \eta \leq 1$} \\
0 & \mbox{otherwise.}
\end{array} \right.
\end{equation}
This is the Haar wavelet, and it is the reason that we used a superscript $H$
on the functions $\phi(x)$ and $\psi(x)$.
As with the mother functions, one can construct a set of
$\psi_{j,l}^{H}(x)$ by dilating and translating eq.(2.10) as
\begin{equation}
\begin{array}{ll}
\psi_{j,l}^{H}(x) & =
   \psi^{H}(2^{j} x/L - l) \\
  \  & = \left\{ \begin{array}{ll}
      1 & \mbox{for $Ll2^{-j} \leq x \leq L(l + 1/2) \  2^{-j}$}\\
      -1 & \mbox{for $L(l+1/2)2^{-j} \leq x \leq L(l+1)\ 2^{-j}$}\\
      0 & \mbox{otherwise.}
\end{array} \right.
\end{array}
\end{equation}
$\psi_{j,l}^{H}(x)$ is called the father function (Meyer 1993)
\footnote{In some of the literature
$\psi_{j,l}^{H}(x)$ is called the mother function, while
$\phi_{j,l}^{H}(x)$ the father function. This is unfortunate but we hope this
confusion will not lead to too many misunderstandings.}

 From eqs.(2.7) and (2.13), we have
\begin{equation}
\begin{array}{ll}
\phi_{j,2l}^{H}(x) &
    = \frac{1}{2} (\phi_{j-1,l}^{H}(x) + \psi_{j-1,l}^{H}(x)),\\
      &   \\
\phi_{j, 2l+1}^{H}(x) &
   = \frac{1}{2}(\phi_{j-1,l}^{H}(x) - \psi_{j-1,l}^{H}(x)).\\
\end{array}
\end{equation}
Eq.(2.4) can be rewritten as
\begin{equation}
\begin{array}{ll}
\epsilon^{J}(x) & =
    \sum_{l=0}^{2^{J-1}-1} \epsilon_{J-1,l} \phi_{J-1,l}^{H}(x)
     + \sum_{l=0}^{2^{J-1}-1} \tilde{\epsilon}_{J-1,l} \psi_{J-1,l}^{H}(x) \\
 \     &   \  \\
\     & = \epsilon^{J-1}(x) +
 \sum_{l=0}^{2^{J-1}-1} \tilde{\epsilon}_{J-1,l} \psi_{J-1,l}^{H}(x)\\
\end{array}
\end{equation}
where we used eqs.(2.8) and (2.9), and
\begin{equation}
\tilde{\epsilon}_{J-1,k}  = \frac {1}{2} (\epsilon_{J,2l} -
   \epsilon_{J,2l+1}).
\end{equation}
The $\tilde{\epsilon}_{j,k}$ are called father function coefficients (FFC).
Thus, the difference between $\epsilon_{J,l}$ and
$\epsilon^{J-1}(x)$ is given by an FFC term as
\begin{equation}
\tilde{\epsilon}^{J-1}(x) \equiv
\sum_{l=0}^{2^{J-1}-1} \tilde{\epsilon}_{J-1,l} \psi_{J-1,l}^{H}(x).
\end{equation}
Eq.(2.15) becomes then
\begin{equation}
\epsilon^{J}(x) = \epsilon^{J-1}(x) +\tilde{\epsilon}^{J-1}(x)
\end{equation}
The term $\tilde{\epsilon}^{J-1}(x)$ contains all information lost during
the window smoothing from order $J$ to $J-1$.  While not immediately
obvious, these two functions, the mother functions and the father functions,
 together form a compactly supported orthogonal bases.

\subsection{Multiresolution analysis}

Eq.(2.18) shows that the $J$-th distribution can be resolved into
a $(J-1)$-th smoothed distribution and a term given by $(J-1)$-th FFCs.
We can repeat this procedure, i.e. resolving the $(J-1)$-th distribution
into $(J-2)$-th smoothed distribution and a term containing the $(J-2)$-th
FFCs, and so
on. The original distribution $\epsilon(x)$ can then be resolved
scale by scale as
\begin{equation}
\begin{array}{ll}
\epsilon^{J}(x) & = \epsilon^{J-1}(x) +\tilde{\epsilon}^{J-1}(x) \\
  & = \epsilon^{J-2}(x) +\tilde{\epsilon}^{J-2}(x) +\tilde{\epsilon}^{J-1}(x)\\
  & ...   \\
  & = \epsilon^{0}(x) + \tilde{\epsilon}^{0}(x) + ...
    +\tilde{\epsilon}^{J-2}(x)    +\tilde{\epsilon}^{J-1}(x)\\
\end{array}
\end{equation}
where
\begin{equation}
\epsilon^{j}(x) = \epsilon^{0}(x) + \tilde{\epsilon}^{0}(x) + ...
    +\tilde{\epsilon}^{j-1}(x) =
   \sum_{l=0}^{2^{j}-1} \epsilon_{j,l} \phi_{j,l}^{H}(x)
\end{equation}
and
\begin{equation}
\tilde{\epsilon}^{j}(x) \equiv
\sum_{l=0}^{2^{j}-1} \tilde{\epsilon}_{j,l} \psi_{j,l}^{H}(x).
\end{equation}

Eq.(2.19) is a wavelet multiscale decomposition (or wavelet
multiresolution analysis) of $\epsilon(x)$. As emphasized before, unlike
the window function decomposition ($\epsilon^{j}(x)$),
the wavelet multiscale decomposition ($\epsilon^{j}(x)$ and
$\tilde{\epsilon}^{j}(x)$) does not lose information. In
other words, one cannot reconstruct $\epsilon(x)$ from
windowed components $\epsilon^{j}(x)$, ($j=0,1,...J-1$), but we are able
to reconstruct the original distribution from the ``difference" functions
$\tilde{\epsilon}^{j}(x)$, ($j=0,1,...J-1$), and $\epsilon^{0}(x)$.
Here $\epsilon^{0}(x)= \epsilon_{0,0}\phi_{0,0}^{H}(x)$, and
$\epsilon_{0,0}$ is simply the mean density of the
distribution $\epsilon(x)$ in the range $[0,L]$. Using (2.1),
we have $\epsilon^{0}(x) =0$.

Moreover, the father functions $\psi_{j,l}^{H}(x)$ are
orthogonal with respect to {\it both} indexes $j$ and $l$, i.e.
\begin{equation}
\int_0^L \psi_{j',l'}^{H}(x)\psi_{j,l}^{H}(x)dx =
\left(\frac{L}{2^j}\right)\delta_{j',j} \delta_{l',l}
\end{equation}
where $\delta_{j',j}$ is the Kronecker delta. For a given $j$,
$\psi_{j,l}^{H}(x)$ are also orthogonal to mother
functions $\phi_{j',l}^{H}(x)$ with $j'\leq j$, i.e.
\begin{equation}
\int_0^L \phi_{j',l'}^{H}(x)\psi_{j,l}^{H}(x)dx = 0,
\hspace{2cm} {\rm if \ \ \ } j' \leq j.
\end{equation}
The FFCs in eq.(2.21) can be found by
\begin{equation}
\tilde{\epsilon}_{j,l} = \frac{2^{j}}{L} \int_0^L
\epsilon(x)\psi_{j,l}^{H} dx.
\end{equation}
The last line of Eq.(2.19) is then
\begin{equation}
\begin{array}{ll}
\epsilon(x)= \epsilon^{J}(x) & = \epsilon^{0}(x) + \tilde{\epsilon}^{0}(x)
+ ...
    +\tilde{\epsilon}^{J-2}(x)    +\tilde{\epsilon}^{J-1}(x)\\
  \  &  \  \\
    & = \sum_{j=0}^{J-1} \sum_{l=0}^{2^{j}-1} \tilde{\epsilon}_{j,l}
 \psi_{j,l}^{H}(x).
\end{array}
\end{equation}
The FFCs provide a complete representation of $\epsilon(x)$ which we will call
the Haar representation.

\subsection{Compactly supported orthogonal bases}

The Haar representation suffers from the drawback that the  $\psi_{j,l}^{H}(x)$
are not localized in Fourier space. As was
mentioned in  \S 2, an adequate space-scale decomposition should be
localized in both physical and scale (Fourier) space. The top-hat
window function (2.6) and the wavelet (2.10) cannot meet this condition,
because they are discontinuous. In the mid-80's to early 90's a great
deal of work was done in trying to find a continuous bases that was
well localized in Fourier space (Daubechies 1988, Meyer 1988, Mallat
1989, Mallat \& Zhong 1990,). Specifically,
Daubechies (1988) constructed several families of wavelets and scaling
functions which are orthogonal, have compact support and are continuous.

In order to construct a compactly supported discrete wavelets basis
the following two recursive equations were involved
 (Daubechies 1992, Meyer 1993).
\begin{equation}
\begin{array}{ll}
\phi(\eta) & = \sum_l a_l \phi(2\eta-l) \\
\psi(\eta) & = \sum_{l} b_{l} \phi(2\eta + l)
\end{array}
\end{equation}
It is easy to show that the Haar scaling (2.6) and wavelet (2.12)
satisfy eq.(2.26) only if the coefficients $a_0=a_1=b_0=-b_1=1$ are
nonvanishing.

Directly integrating  the first equation in (2.26) it follows
that
\begin{equation}
\sum_l a_l= 2.
\end{equation}
Requiring orthonormality for $\phi(x)$ with respect to discrete integer
translations, i.e.
\begin{equation}
\int_{-\infty}^{\infty} \phi(\eta-m) \phi(\eta) d\eta = \delta_{m,0},
\end{equation}
we have that
\begin{equation}
\sum_l a_l a_l+2m = 2 \delta_{0,m}.
\end{equation}
The wavelet $\psi(\eta)$ has to qualify as a ``difference" function, i.e.
it is admissible. We have then
\begin{equation}
\int_{-\infty}^{+\infty}\psi(\eta) d\eta = 0,
\end{equation}
so we need
\begin{equation}
\sum_l b_l = 0
\end{equation}
The multiresolution analysis requires that
\begin{equation}
\int_{-\infty}^{\infty} \psi(\eta)\phi(\eta -l) d\eta = 0.
\end{equation}
So one has
\begin{equation}
b_l=(-1)^la_{1-l}.
\end{equation}
After the Haar wavelet, the simplest solution of the recursive equations
(2.26) with conditions (2.27), (2.29) is
\begin{equation}
a_0=(1+\sqrt 3)/4, \ a_1=(3+\sqrt 3)/4, \
a_2=(3-\sqrt 3)/4, \ a_3=(1-\sqrt 3)/4.
\end{equation}
This is called the Daubechies 4 wavelet (D4) and
is plotted in Figure 2. In general, the more
coefficients $a_l$ that are nonvanishing, the wider the compact
support is, while the wavelet itself becomes smoother
(Daubechies 1992; Chui 1992; Kaiser, 1994)

\begin{figure}[htb]
\begin{center}
\epsfig{file=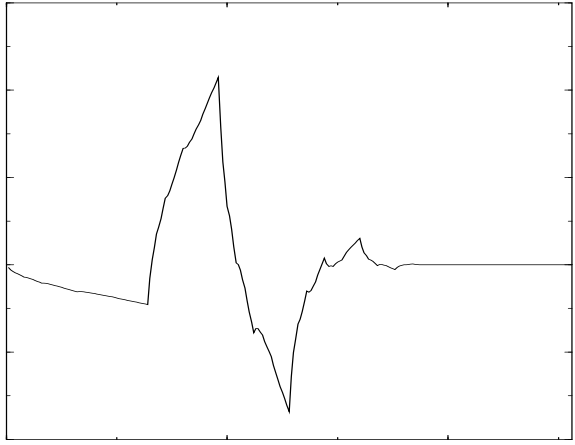,height=3in,width=4in}
   \label{fig2}
\vspace{-10mm}
\caption{The Daubechies 4 wavelet, $\psi(\eta)$, determined by (2.34).
The wavelet is plotted for presentation purposes only, so the
units are arbitrary.}
\end{center}
\end{figure}

\subsection{DWT decomposition}

To conduct a DWT analysis, one first constructs the DWT bases
by dilation and translation of $\phi(x)$ and $\psi(x)$ as
\begin{equation}
\phi_{j,l}(x) = \left( \frac{2^j}{L} \right )^{1/2} \phi(2^jx/L - l)
\end{equation}
and
\begin{equation}
\psi_{j,l}(x) =\left( \frac{2^j}{L} \right )^{1/2} \psi(2^jx/L-l),
\end{equation}
where $\psi_{j,l}$ and $\phi_{j,l}$ with integer $j$ and $l$
are the father functions and mother functions, respectively.
Different from eqs.(2.7) and (2.13), eqs.(2.35) and (2.36) include a
normalization factor $(2^j/L)^{1/2}$. The set of $\psi_{j,l}$ and
 $\phi_{0,m}(x)$ with $ 0 \leq j < \infty$
and $- \infty < l, m < \infty$ form a complete, orthonormal basis in the
space of functions with period length $L$.

To subject a finite sample in region $L$ to a DWT expansion, we
introduce an auxiliary density distribution $\rho(x)$ which is an
$L$ periodic function defined on space
$- \infty < x < \infty$. Using the complete, orthonormal basis,
$\psi_{j,l}$ and $\phi_{j,m}(x)$, the density distribution $\rho(x)$
can be expanded as
\begin{equation}
\rho(x) = \sum_{m=-\infty}^{\infty} c_{0,m}\phi_{0,m}(x) +
 \sum_{j=0}^{\infty} \sum_{l= - \infty}^{\infty}
 \tilde{c}_{j,l} \psi_{j,l}(x)
\end{equation}
where the coefficients $c_{0,m}$ and $\tilde{c}_{j,l}$  are
calculated by the inner products
\begin{equation}
c_{0,m}=\int_{-\infty}^{\infty} \rho(x) \phi_{0,m}(x) dx,
\end{equation}
and
\begin{equation}
\tilde{c}_{j,l}=\int_{-\infty}^{\infty} \rho(x) \psi_{j,l}(x)dx.
\end{equation}
Because all bases functions
$\int_\infty^\infty \psi_{j,l}(x)dx =0$ [eq.(2.30)], eq.(2.37) can be
rewritten as
\begin{equation}
\rho(x) = \sum_{m=-\infty}^{\infty} c_{0,m}\phi_{0,m}(x) +
\bar{\rho} \sum_{j=0}^{\infty} \sum_{l= - \infty}^{\infty}
 \tilde{\epsilon}_{j,l} \psi_{j,l}(x)
\end{equation}
where $\bar\rho$ is the mean density, and
\begin{equation}
\tilde{\epsilon}_{j,l}=\int_{-\infty}^{\infty} \epsilon(x) \psi_{j,l}(x)dx
\end{equation}
where $\epsilon(x)= (\rho(x)-\bar{\rho})/\bar{\rho}$ is the density
contrast as eq.(2.1).

By definition, $\rho(x)=\rho(x+mL)$ for integers $m$,
eq.(2.38) can be rewritten as
\begin{equation}
c_{0,m}=\int_{-\infty}^{\infty} \rho(x+mL) \phi_{0,m}(x) dx.
\end{equation}
Using eq.(2.35), we have then
\begin{eqnarray*}
c_{0,m} & = & \int_{-\infty}^{\infty} \rho(x+mL) L^{-1/2} \phi(x/L - m) dx \\
    & = & \int_{-\infty}^{\infty} \rho(x') L^{-1/2} \phi(x'/L ) dx'
\end{eqnarray*}
\begin{equation}
\ \ \  = \ \int_{-\infty}^{\infty} \rho(x') \phi_{0,0}(x') dx' = c_{0,0}
\end{equation}
where $x'=x+mL$. The coefficients $c_{0,m}$ are independent of $m$.
Using the property of ``partition of unity" of the scaling
function (Daubechies 1992)
\begin{equation}
\sum_{m=-\infty}^{\infty}\phi(\eta +m)=1,
\end{equation}
from eq.(2.43) one has
\begin{eqnarray*}
c_{0,0} & = & \int_{-\infty}^{\infty} \rho(x) L^{-1/2} \phi(x/L ) dx \\
    & = & \sum_{-\infty}^{\infty}\int_0^L \rho(x+mL) L^{-1/2} \phi(x/L +m) dx
\\
    & = & L^{-1/2} \int_0^L \rho(x)\sum_{-\infty}^{\infty} \phi(x/L +m) dx
\end{eqnarray*}
\begin{equation}
\ \   = \ L^{-1/2} \int_0^L \rho(x) dx. \  \  \  \
\end{equation}
The first term in the expansion (2.40) becomes
\begin{equation}
\sum_{m=-\infty}^{\infty} c_{0,m}\phi_{0,m}(x) =
c_{0,0} L^{-1/2} \sum_{m=-\infty}^{\infty} \phi(x/L +m)
=L^{-1}\int_0^L \rho(x)dx.
\end{equation}
This term is the mean density $\bar \rho$. From eq.(2.40)
we have finally
\begin{equation}
\epsilon(x) = \frac{\rho(x)-\bar\rho}{\bar\rho}
   = \sum_{j=0}^{\infty} \sum_{l= - \infty}^{\infty}
  \tilde{\epsilon}_{j,l} \psi_{j,l}(x)
\end{equation}
where the father function coefficients $\tilde{\epsilon}_{j,l}$ are
given by eq.(2.41). Eq.(2.47) has the same terms as eq.(2.25),
and is a multiresolution analysis with respect to
the orthogonal bases $\psi_{j,l}(x)$.

While the multiresolution analysis eq.(2.19) still holds,  eqs.(2.20) and
(2.21) should be replaced by
\begin{equation}
\epsilon^{j}(x) = \tilde{\epsilon}^{0}(x) + \tilde{\epsilon}^{1}(x) ...
    +\tilde{\epsilon}^{j-1}(x) =
   \sum_{l=0}^{2^{j}-1} \epsilon_{j,l} \phi_{j,l}(x)
\end{equation}
and
\begin{equation}
\tilde{\epsilon}^{j}(x) \equiv
\sum_{l=0}^{2^{j}-1} \tilde{\epsilon}_{j,l} \psi_{j,l}(x).
\end{equation}

\subsection{Relationship between Fourier and DWT expansion}

In terms of the Fourier transform, $\epsilon(x)$ is expressed as
\begin{equation}
\epsilon(x)=\sum_{n = - \infty}^{\infty} \epsilon_n e^{i2\pi nx/L}
\end{equation}
with the coefficients computed by
\begin{equation}
\epsilon_n= \frac{1}{L}\int_0^{L} \epsilon(x)e^{-i2\pi nx/L}dx.
\end{equation}

Since both the DWT and Fourier bases sets are complete, a function may be
represented by either bases and there is thus a relationship between the
FFCs and the Fourier coefficients.
Substituting expansion (2.50) into eq.(2.41), yields
\begin{equation}
\tilde{\epsilon}_{j,l}=\sum_{n= -\infty}^{\infty} \epsilon_n
\int_{-\infty}^{\infty} e^{i2\pi nx/L} \psi_{j,l}(x)dx
= \sum_{n= -\infty}^{\infty} \epsilon_n \hat{\psi}_{j,l}(-n)
\end{equation}
where $\hat{\psi}_{j,l}(n)$ is the Fourier transform of
$\psi_{j,l}(x)$, i.e.
\begin{equation}
\hat{\psi}_{j,l}(n) = \int_{-\infty}^{\infty}
\psi_{j,l}(x)e^{-i2\pi nx/L}dx.
\end{equation}
Using eq.(2.36), eq.(2.52) gives
\begin{equation}
\tilde{\epsilon}_{j,l} = \sum_{n = - \infty}^{\infty}
\left(\frac{ 2^{j}}{L}\right )^{1/2} \epsilon_n \int_{-\infty}^{\infty}
 e^{i2\pi nx/L} \psi(2^jx/L - l) dx \ .
\end{equation}
Defining variable $\eta=2^jx/L-l$, one finds
\begin{equation}
\tilde{\epsilon}_{j,l} = \sum_{n = - \infty}^{\infty}
\left(\frac{ 2^{j}}{L}\right )^{-1/2} \epsilon_n e^{i2\pi nl/2^j}
\int_{-\infty}^{\infty}  e^{i2\pi n \eta/2^j} \psi(\eta) d \eta
\end{equation}
or
\begin{equation}
\tilde{\epsilon}_{j,l} = \sum_{n = - \infty}^{\infty}
\left(\frac{ 2^{j}}{L}\right )^{-1/2} \epsilon_n
\hat{\psi}(-n/2^j)  e^{i2\pi nl/2^j}
\end{equation}
where $\hat{\psi}(n)$ is the Fourier transform of the wavelet
$\psi(\eta)$
\begin{equation}
\hat{\psi}(n)=\int_{-\infty}^{\infty} \psi(\eta) e^{-i2\pi n\eta}d\eta.
\end{equation}
Eq.(2.56) is the expression of the FFCs in terms of the Fourier coefficients.

 From expansions (2.51) and (2.49), one can also express
the Fourier coefficient, $\epsilon_n$, in terms of FFCs as
(see Appendix A)
\begin{equation}
\epsilon_n = \frac{1}{L}\sum_{j=0}^{\infty} \sum_{l=0}^{2^j-1}
\tilde{\epsilon}_{j,l} \hat{\psi}_{j,l}(n), \ \ \ \ \ \
\ \ \ n \neq 0
\end{equation}
or
\begin{equation}
\epsilon_n =  \sum_{j=0}^{\infty}  \sum_{l=0}^{2^j-1}
\left (\frac{1}{2^{j}L}\right )^{1/2}
\tilde{\epsilon}_{j,l} e^{-i2\pi nl/2^j} \hat{\psi}(n/2^j), \ \ \ \ \ \
\ \ \ n \neq 0.
\end{equation}
Eq.(2.58) and (2.59) show how the Fourier coefficients are determined by a
DWT analysis.

\subsection{Comparison with the Fourier transform}

The difference between the Fourier transform and the DWT can easily
be seen in phase space $(x,k)$. According to the uncertainty principle,
each mode of a complete, orthogonal bases set corresponds to a "element"
with size $\Delta x$ and $\Delta k$, and area $\Delta x \Delta k \sim 2
\pi$ in phase space. For the Fourier transform,
the "elements" are taken to be $\Delta k=0$ and $\Delta x = \infty$, while
for the DWT both $\Delta k$ and $\Delta x$ are
 finite, and $\Delta x \Delta k \simeq 2 \pi$ (Figure 3).

\begin{figure}[htb]
\begin{center}
\epsfig{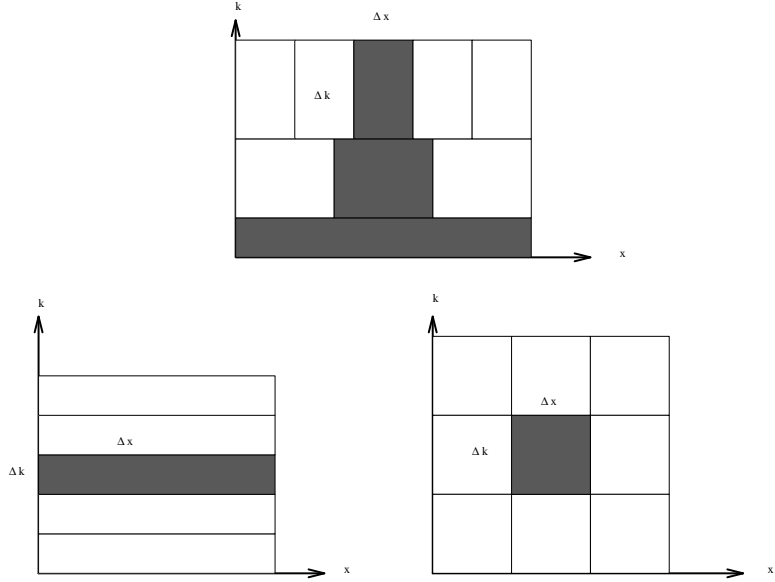}
   \label{fig3}
\caption{``Element" of different transforms in phase space  $(x,k)$.
A. DWT, both $\Delta k$ and $\Delta x$ are finite,
and $\Delta x \Delta k \simeq 2 \pi$.  B. Fourier transform, the "elements"
are $\Delta k=0$ and $\Delta x = \infty$. C. Gabor transform,
here $\Delta k= \Delta x$ = constant, regardless the scale.}
\end{center}
\end{figure}

The representation by the Fourier bases, i.e., the trigonometric functions,
is delocalized ($\Delta x = \infty$). The DWT resolves the distribution
$\epsilon(x)$ simultaneously in terms of its standard variable (say space)
and its conjugate counterpart in Fourier space (wavenumber or scale) up to
the limit of uncertainty principle.

This doesn't mean that the Fourier transform loses information about
$\epsilon(x)$ but rather that the information on the position is spread out.
For instance, let us consider a distribution that contains a few clumps of
scale $d$. The positions of the clumps are related to the phases of all the
Fourier coefficients $k < 2\pi/d$. There is no way to
find the positions of the clumps from a finite number of Fourier coefficients.
The only solution would be to reconstruct $\epsilon(x)$ from {\it all} the
Fourier coefficients. If some of the ''clumps" are due to experimental
errors, we will not be able to filter them out because they have affected
all the Fourier coefficients.

On the other hand, the DWT keeps the locality present in the distribution
and allows for the local reconstruction of a distribution. It is then possible
to reconstruct only a portion of it. There is a relationship between the
local behavior of a distribution and the local behavior of its FFCs.
For instance, if a distribution $\epsilon(x)$ is locally smooth, the
corresponding FFCs will remain small, and if $\epsilon(x)$ contains a clump,
then in its vicinity the FFCs amplitude will increase drastically.
To reconstruct a portion of the distribution, it is only
necessary to consider the FFCs belonging to the corresponding subdomain
of the wavelet space $(j,l)$. If the FFCs are occasionally subject to
errors, this will only affect the reconstructed distribution locally
near the flawed positions. Furthermore, the Fourier transform is also
particularly sensitive to phase errors due to the alternating character of the
trigonometric series. This is not the case for the DWT.

Early approaches to finding information on the locations in the Fourier
transform schemes were given by the Wigner function in quantum mechanics and
the Gabor transform. The difference between the Gabor transform
and the DWT can also seen in Figure 3. The orthogonal basis of the DWT
are obtained by (space) translation and (scale) dilation of one wavelet.
The DWT transform gives very good spatial resolution on small scales, and very
good scale resolution on large scales. Gabor's windowed Fourier transform
is based on a family of trigonometric functions exhibiting
increasingly many oscillations in a window of constant size.  In this
case the spatial resolution on  small scales and the range on large
scales are limited by the size of the window.

The relation between the discrete and the continuous
Fourier transforms when $\epsilon(x)$ is viewed as a continuous distribution
sampled on an interval $\Delta$ is
$\hat{\epsilon}(n) = \Delta \hat{\epsilon}_n$
where $\hat{\epsilon}(n)$ is the usual continuous Fourier transform and
$\hat{\epsilon}_n$ is the discrete Fourier transform.
  However, no such
relationship exists between the discrete wavelet transform, such as the
D4 wavelet, and the
continuous wavelet transform (CWT). The
D4 wavelet has no continuous analog.  Each type of complete, orthogonal
wavelet bases must be constructed from scratch.  In general, CWTs form
an over complete bases, i.e. they are highly redundant.  As a consequence,
CWT coefficients of a random sample show A correlation that is not in
the sample itself but is given by the over complete bases.  Since
the DWT is complete and orthogonal, this is not a problem in using the DWT.
The point is that the difference between the DWT and the CWT is essential, and
by construction, these two bases are very different.

The DWT is not intended to replace the Fourier transform,
which remains very appropriate in the study of all topics where there is no
need for local information.

\setcounter{enumi}{3}
\setcounter{equation}{0}

\section{A DWT Estimation of the Probability Distribution Function}

\subsection{One-point distribution of FFCs}

As has been emphasized in \S 1, the statistical features of a non-linearly
evolved density field can not be completely described by the power spectrum or
the two-point correlation function. For a complete description of salient
statistical features, the probability
distribution functions (PDF) of the density field are required.

There is, however, a very real problem in determining the PDF due mainly to the
central limit theorem of random fields (Adler 1981).
If the universe consists of a large number of
randomly distributed clumps with a non-Gaussian one-point function, eq.(2.51)
shows that for large $L$ the Fourier amplitudes, $\epsilon_n$, are given by
a superposition of a large number of non-Gaussian clumps. According to the
central limit theorem, the distribution of $\epsilon_n$ will be Gaussian
when the total number of clumps is large. Thus, in general, the statistical
features of the clumps can not be seen from the one-point distribution of
the Fourier modes, $\epsilon_n$, even if the PDF function of clumps is
highly non-Gaussian.  If the clumps are not distributed independently,
but are correlated, the central limit theorem still holds if the two-point
correlation function of the clumps approaches zero sufficiently fast (Fan
\& Bardeen 1995).

On the other hand, the father functions, $\psi_{j,l}(x)$, are localized.
If the scale of the clump is $d$, eq.(2.41) shows that the FFC,
$\tilde{\epsilon}_{j,l}$, with $j=\log_2(L/d)$,  is determined only by
the density field in a range containing no more than one clump.
That is, for scale $j$ the FFCs are not given by a superposition of a
large number of the clumps, but determined by at most one of them. Thus, the
one point distribution of the FFC, $\tilde{\epsilon}_{j,l}$, avoids the
restriction of the central limit theorem, and is able to detect the PDF
related to the clumps, regardless the total number of the clumps
in the sample being considered.

This point can also be shown from the orthonormal bases being used for the
expansion of the density field. A key condition needed for the
central limit theorem to hold is that the modulus of the bases be less
than $C/\sqrt L$, where $L$ is the size of the sample and $C$ is a constant
(Ivanonv \& Leonenko 1989). Obviously, all Fourier-related orthonormal
bases satisfy this condition because the Fourier orthonormal bases
in 1-dimension are such that $(1/\sqrt L) |\sin kx| < C/\sqrt L$
and $(1/\sqrt L)|\cos kx| < C/\sqrt L$, and $C$ is independent of
coordinates in both physical space $x$ and scale space $k$. On the other
hand, the father functions (2.36) have
\begin{equation}
|\psi_{j,l}(x)| \sim \left( \frac{ 2^{j}}{L}\right )^{1/2} {\rm O}(1)
\end{equation}
because the magnitude of the basic wavelet $\psi(x)$ is of the order
1 [eq.(3.26)]. The condition $|\psi_{j,l}(x)| < C/ \sqrt{L}$, will no
longer hold for a constant $C$ independent of scale variable $j$.

\subsection{``Ensemble" of FFCs}

In cosmology, no ensemble of cosmic density fields exists, and at most
only one realization, i.e. the observed density distribution, is available.
In order to have reasonable statistics, the cosmic density field is usually
assumed to be ergodic: the average over an ensemble is equal to the spatial
average taken over one realization. It is sometimes called the ``fair sample
hypothesis" in LSS study (Peebles 1980). A homogeneous Gaussian
field with continuous spectrum is certainly ergodic (Adler 1981). In
some non-Gaussian cases, such as homogeneous and isotropic turbulence
(Vanmarke, 1983), ergodicity also approximately holds. Roughly, the
ergodic hypothesis is reasonable if spatial correlations are decreasing
sufficiently rapidly with increasing separation.  The volumes separated
with distances larger than the correlation length are approximately
statistically independent. Even for one realization of
$\epsilon(x)$, FFCs at different locations $l$ can be treated as results
from statistically independent realizations. Thus, the values of
$\epsilon_{j,l}$ on different $l$ form an ensemble of the FFCs on scale $j$.

This result can be more clearly seen from eq.(2.56). For many wavelets,
their Fourier transforms $\hat{\psi}(n)$ are non-zero (localized) in two
narrow ranges centered at
$n =\pm n_p$. For instance, the Battle-Lemari\'{e} wavelet $n_p=\pm 1$,
and the Daubechies 4 wavelet $n_p \sim \pm 1.9$.
The sum over $n$ in eq.(2.56) is only taken over the two ranges
$(n_p - 0.5 \Delta n_p)2^j \leq n \leq (n_p + 0.5\Delta n_p)2^j$ and
$-(n_p + 0.5 \Delta n_p )2^j \leq n \leq -(n_p - 0.5 \Delta n_p )2^j$,
where $\Delta n_p $ is the width of the non-zero ranges of $\hat{\psi}(n)$.
Eq.(2.56) can then be approximately rewritten as
\begin{eqnarray*}
\tilde{\epsilon}_{j,l} & \simeq & \left(\frac{L}{2^{j}} \right)^{1/2}
  2\sum_{n=(n_p - 0.5\Delta n_p)2^j}^{(n_p + 0.5\Delta n_p)2^j}
  {\rm Re}\{ \hat{\psi}(n_p)\epsilon_{n} e^{i2\pi nl/2^j} \}
\end{eqnarray*}
 \begin{equation}
\ \ \ \ \ \ \ \ \ \ \simeq \left(\frac{L}{2^{j}}\right)^{1/2}
|\hat{\psi}(n_p)|
2\sum_{n=(n_p - 0.5\Delta n_p)2^j}^{(n_p + 0.5\Delta n_p)2^j}
 |\epsilon_{n}|\cos (\theta_{\psi}+\theta_{n}+2\pi nl/2^j)
\end{equation}
where we have used $\hat{\psi}(-n_p)=\hat{\psi}^*(n_p)$ and
$\epsilon_{-n}= \epsilon^*_{n}$, because both $\psi(x)$ and $\epsilon(x)$ are
real. $\theta_{\psi}$, $\theta_n$ in eq.(9) are the phases of
$\hat{\psi}(n_p)$ and $ \epsilon_n$, respectively.

As we pointed out in \S 3.1, $\epsilon_n$ is Gaussian even when the clumps
are non-Gaussian. For a homogeneous random field, the phase of $\epsilon_n$,
i.e. $\theta_n$, should be uniformly randomly distributed and
from eq.(3.2), the probability  distribution of $\tilde{\epsilon}_{j,l}$ is
independent of $l$. Thus, each  FFC is a realization
of the $l-independent$ stochastic variable $\tilde{\epsilon}_{j,l}$. The FFCs,
$\tilde{\epsilon}_{j,l}$, on scale $j$ form an ensemble with $2^j$
realizations.
The statistics with respect to the one-point distribution of FFCs
$\tilde{\epsilon}_{j,l}$ should be equal to the results
of the ensemble statistics. The goodness of this estimation can be measured by
the Large Number Theorem, that is, the relative error is about $1/\sqrt{2^j}$.
Simply stated, when the ``fair sample hypothesis" holds, the one-point
distribution
of FFCs from (observed) one-realization will be a fair estimate of the PDF
of the cosmic density field.

\subsection{Scale mixing}

The PDF is sometimes measured by the count in cell (CIC) method. The CIC
detects the one-point distribution of the density field in given
cubical cells with side $d$ or Gaussian spheres with radius $R_G$. It is
generally believed that the CIC one-point distribution of window $d$ is
dominated by the density fluctuations on scale $\sim 2d$ or $R_G$.

Actually, the CIC analysis is essentially the same as the MFC. The window
function of the CIC corresponds to the mother function, and the count to
the amplitude of the MFCs. The one-point distribution given by
the CIC have similar properties as the MFC one-point distribution. A
problem with the MFC one-point distributions is scale mixing. Even though the
mother
functions of the DWT, $\phi_{j,l}(x)$, are localized in spatial space they
are not orthogonal with respect to the scale index $j$, i.e. not localized
in Fourier space. The MFCs, $\phi_{j,l}(x)$, are dependent on perturbations
on all scales larger than $L/2^j$. Thus, if the clumps are multiscaled, the
MFCs will also be Gaussian as required by the central limit theorem and the
MFC one-point distribution may miss the clumps. Similarly, the CIC is
scale mixed, and not suitable for studying the
scale dependence (or spectrum) of various statistical measures.

There are more problem related to the cubic cell CIC. The cubic cell window is
just the Haar wavelet or Daubechies 2 (D2) [see eq.(2.5)]. It is very well
known that among Daubechies wavelets, only D2 is not localized
in Fourier space because the Fourier transform of its wavelet (2.12) is
\begin{equation}
\hat{\psi}^H(n)=\frac{2}{\pi n}[\sin(\pi n) - i \cos(\pi n)]
\sin^2 (\pi n/2)
\end{equation}
When $n \ll 1$, $\hat{\psi}^H(n) \sim -i(\pi/2)n$. Therefore, eq.(2.56) gives
\begin{equation}
\tilde{\epsilon}_{j,l}  = \sum_{n < 2^j}
\left(\frac{ 2^{j}}{L}\right )^{-1/2} \epsilon_n
\frac{i\pi An}{2^{j+1}} e^{i2\pi nl/2^j} + {\rm terms} \ \  n \geq 2^j
\end{equation}
Eq.(3.4) shows that the large scale (small $n$) perturbations will
significantly contribute to, and even dominate the $\epsilon_{j,l}$ if
$\lim_{n \rightarrow 0} n\epsilon$ is larger than a non-zero constant.
So even the FFC one-point distributions of the D2 wavelet are not
a good way of measuring the PDF.

\subsection{Spectrum of FFC cumulants}

Since FFC one-point distributions effectively allow for scale
decomposition, one
can easily calculate the spectrum of the FFC moments or cumulants on any
order as follows.

The second order cumulant is given by
\begin{equation}
\sigma^2_j \equiv \frac{1}{2^j} \sum_{l=0}^{2^j-1}
(\tilde{\epsilon}_{j,l} - \overline{\tilde{\epsilon}_{j,l}})^2,
\end{equation}
where $\overline{\tilde{\epsilon}_{j,l}}$ is the average of
$\tilde{\epsilon}_{j,l}$ over $l$. $\sigma^2_j $ is, in fact, the power
spectrum with respect to the modes of DWT (see, \S 4.1).

The third and fourth orders are
\begin{equation}
C^3_j \equiv \frac{1}{2^j} \sum_{l=0}^{2^j-1}
(\tilde{\epsilon}_{j,l} - \overline{\tilde{\epsilon}_{j,l}})^3,
\end{equation}
\begin{equation}
C^4_j \equiv \frac{1}{2^j} \sum_{l=0}^{2^j-1}
(\tilde{\epsilon}_{j,l} - \overline{\tilde{\epsilon}_{j,l}})^4 - 3\sigma^4_j,
\end{equation}
$C^3$ and $C^4$ will measure the spectra of skewness and kurtosis,
respectively (see, \S 5.1).

Generally, the spectrum of $n$-th order moment is defined as
\begin{equation}
C^n_j \equiv \frac{1}{2^j} \sum_{l=0}^{2^j-1}
(\tilde{\epsilon}_{j,l} - \overline{\tilde{\epsilon}_{j,l}})^n,
\end{equation}
The definitions (3.5) - (3.8) show that the numerical work of calculating
higher order moments (cumulant) is not any more difficult than calculating
the second order moments.
Generally, the calculation of third and higher order correlations of large
scale structure samples is very strenuous work. But the numerical work
involved in calculating the DWT is not more difficult than the FFT, and can
be faster.  The FFT requires $\sim$ N Log N calculations, while the DWT,
using a ``pyramid" scheme, requires only order N calculations (Press et al.
1992).

\setcounter{enumi}{4}
\setcounter{equation}{0}

\section{Local power spectrum}

\subsection{Power Spectrum with respect to DWT basis}

For the Fourier expansion (2.50) and (2.51)  Parseval's theorem is
\begin{equation}
\frac{1}{L} \int_0^L |\epsilon(x)|^2 dx =
   \sum_{n= -\infty}^{\infty} |\epsilon_n|^2,
\end{equation}
which shows that the perturbations can be decomposed into domains, $n$,
by the orthonormal Fourier basis functions. The power spectrum of
perturbations on length scale $L/n$ is then defined as
\begin{equation}
P(n)= |\epsilon_n|^2.
\end{equation}

Similarly, the Parseval theorem for the expansion
(2.49) can be shown to be (see Appendix B)
\begin{equation}
\frac{1}{L}\int_0^L |\epsilon(x)|^2 dx = \sum_{j= 0}^{\infty}
\frac{1}{L}\sum_{l=0}^{2^j-1} |\tilde{\epsilon}_{j,l}|^2.
\end{equation}
Comparing eqs.(3.3) and (3.1), one can relate the term
$\sum_{l=0}^{2^j-1}|\tilde{\epsilon}_{j,l}|^2/L$ to the
power of perturbations on length scale $L/2^j$, and the term
$|\tilde{\epsilon}_{j,l}|^2/L$ to the power of the perturbation
on scale $L/2^j$ at position $lL/2^j$.
The spectrum with respect to the DWT bases can be defined as
\begin{equation}
P_j =\frac{1}{L}\sum_{l=0}^{2^j-1} |\tilde{\epsilon}_{j,l}|^2.
\end{equation}

The DWT spectrum should be defined as the variance
of the FFCs, i.e.,
\begin{equation}
P^{var}_j=\frac{1}{L}\sum_{l=0}^{2^j-1}
 (\overline{\tilde{\epsilon}_{j,l}} - \tilde{\epsilon}_{j,l})^2.
\end{equation}
Because the mean of the FFCs, $\overline{\tilde{\epsilon}_{j,l}}$, over
$l$ is zero [eq.(2.56)], $P_j$ should be equal to $P^{var}_j$.
Comparing with eq.(3.5), we have $\sigma^2_j=(L/2^j) P^{var}_j$ .

\subsection{Relationship between Fourier and DWT Spectra}

The relationship between the spectra of the Fourier expansion (4.2) and
the DWT (4.4) or (4.5) can be found from eq.(3.2), which shows that the
FFCs on scale $j$ are mainly determined by the Fourier components
$\epsilon_n$, with $n$ centered at
\begin{equation}
n=\pm n_p 2^j,
\end{equation}
where $|n_p|$ are the positions of the peaks of $\hat\psi(n)$.
Thus, from eqs.(4.4), (4.5) and (4.6), we have
\begin{equation}
P(n)_j \simeq
\frac{1}{2^{j+1}\Delta n_p}|\hat{\psi}(n_p)|^{-2}P^{var}_j
\end{equation}
where
$P(n)_j$ is the average of Fourier spectrum on the scale $j$ given by
\begin{equation}
P(n)_j = \frac{1}{2^{j}\Delta n_p}
\sum_{n=(n_p - 0.5\Delta n_p)2^j}^{(n_p + 0.5\Delta n_p)2^j}P(n).
\end{equation}
Eqs. (4.7) and (4.8) provide the basic way of detecting the Fourier power
spectrum by a DWT analysis.

 From eqs.(4.8) and (4.6), one has
\begin{equation}
\log P(k)_j = \log P_j - (\log 2) j + A,
\end{equation}
and
\begin{equation}
\log k= (\log 2)j - \log L/2\pi + B.
\end{equation}
The factors $A$ and $B$ are given by
\begin{equation}
A = - \log(2\Delta n_p|\hat{\psi}(n_p)|^{2})
\end{equation}
and
\begin{equation}
B= \log n_p.
\end{equation}
The constants $A$ and $B$ depend on the basic wavelet
$\psi(\eta)$ being used in the DWT analysis. In the case
of the D4 wavelet, $A= 0.602$, and $B= 0.270.$

Eqs.(4.9) - (4.12) provide the way of directly transferring a DWT spectrum
$P_j$ or $P^{var}_j$ into the corresponding Fourier spectrum $P(k)_j$ and
{\it vice versa}.

\subsection{Finite size and complex geometry of samples}

It is well known that a difficulty in spectrum detection comes from
the finite size of the sample which leads to uncertainty in the mean
density of the objects being considered. The classical spectrum
estimator, i.e., the Fourier transform of the two-point correlation function
depends essentially on the mean density $\overline\rho$. A
two-point correlation analysis cannot detect any correlations with amplitude
comparable to the uncertainty of the mean density. If we determine the spectrum
via the two-point correlation function, the uncertainty in $\overline{\rho}$
leads to uncertainties on all scales in which the correlation amplitude
is comparable to the uncertainty in the mean density. The problem of the
uncertainty in the mean density is more severe for the study of high redshift
objects because the mean density of these objects generally is
redshift-dependent.

The CIC or Fourier transform on a finite domain have been used  to avoid
this difficulty. The CIC detects the variance of density
fluctuations in windows of a cubical cell with side $l$ or Gaussian
sphere with radius $R_G$. The behavior of the perturbations on scales larger
than the size of a sample is assumed not to play an important role.
This reduces the uncertainties caused by a poor knowledge of long wavelength
perturbations and by the finite size of the observational samples. It is
believed that the variance in cell $l$ is mainly dominated by the
perturbation on scale $\sim 2l$ or $R_G$. Therefore, the variances
are considered to be a measure of the power spectrum on scale $l$
(Efstathiou et al 1990, Kaiser \& Peacock 1991). Additionally, the Fourier
transform on a
finite domain (the Gabor transform, for instance) can also avoid the
difficulty of the infinity of $x$.

As was mentioned in \S 3.3, the problem with the CIC statistic is that
the variances obtained from the decomposition of cells with different size
$d$ are not independent. While it is still possible to
reconstruct the power spectrum by $\sigma^2(l)$, the scale mixing leads to
uncertainty. The scale mixing becomes a serious problem when the power law
spectrum has a negative index. Moreover, the resulting errors are not easy
to interpret because the errors for different $d$ are also not independent.

The DWT provides a method to solve this problem. Actually, this problem is the
same as trying to detect a  {\it local} (finite range) spectrum.
One of the motivations of developing DWT was to measure local spectra (Yamada
and Ohkitani 1991, Farge 1992). Because the FFCs are localized,  one can
calculate
local power spectrum using the FFCs related to the finite range being
considered. The FFCs are determined by the difference of the MFCs in a
localized neighborhood, i.e., by measuring the differences between the
{\it local} mean densities. As a consequence, the mean density over length
scales larger than the sample's size is not necessary in calculating the FFCs
on scales equal to or less than the size of data. The influence of
the uncertainty of the mean density is significantly reduced by the FFC
spectrum detection.

To avoid the difficulty of finite sized samples, specific boundary conditions
are sometimes selected. For instance, N-body simulations always assume a
periodic
condition. Yet, the choice of boundary conditions may affect the spectrum
detection. Again, father functions $\psi_{j,l}(x)$ are compactly supported,
and the FFCs that are uncertain due to the boundary conditions are only the
coefficients at the  boundary, $\tilde\psi_{j,l_1}$ and $\tilde\psi_{j,l_2}$,
where $l_1$ and $l_2$ are the positions of the 1-D boundary. The uncertainty
due to the choice of boundary conditions will also be suppressed by the wavelet
SSD. We illustrate this point by using different boundary conditions to
find the spectrum. Figure 4 shows the local spectrum of sample in a finite
area $L$ with 512 bins, with boundary conditions  A) periodic outside this
range; B) zero outside this range. Excluding a small effect on the largest
scale (i.e. the scale of the size of the sample), all the local spectra remain
unchanged regardless of which boundary conditions is used.
\begin{figure}[htb]
\begin{center}
\epsfig{file=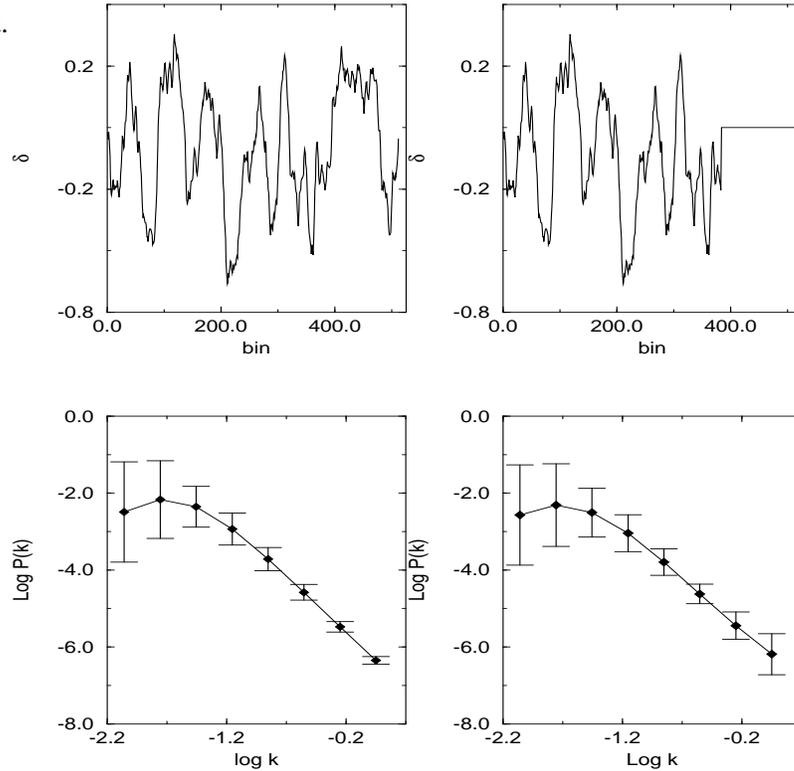,height=4in,width=4in}
   \label{fig4}
\vspace{-8mm}
\caption{Density distributions generated from spectrum
$P(k)=k/(1+10^5k^4)$ in a range of $L=512$ bins, where $k=2\pi n/L$,
and $n$ is integer. The boundary conditions outside the 512 bin
area are taken to be A) periodic, B) zero.
C.) and D.) are the spectrum reconstruction for A.) and B.)
respectively.}
\end{center}
\end{figure}

More precisely, the effect of boundary conditions on the FFCs should be
described by the so-called ``influence cone" which consists of the
spatial support of all dilated father functions. For instance, if
$\psi_{jl}(x)$ is well-localized in the space interval $\Delta x$ for
$j=0$, the influence cone centered at $x_0$ will be defined by
$x \in [x_0 - (\Delta x/2^{j+1}), x_0 + (\Delta x/2^{j+1})]$. Namely,
the FFCs corresponding to the positions with distance larger than
$(\Delta x/2^{j+1})$ from the boundary will not be affected by the selection
of boundary conditions.

This result is also useful in solving the problem of the complex geometry
of observational data. For instance,  samples of Ly$\alpha$ forests in QSO
absorption spectra cover different spatial (redshift) ranges for different
QSO's. Any statistic of a sample compiled from many QSO spectra needs,
at the very least, a complicated weighting scheme. Moreover, the number density
of Ly$\alpha$ forest lines depends on redshift, and therefore, it is very
difficult to find a proper weighting scheme for methods which depends
essentially
on a good measure of the mean density of the sample. Despite the fact that
samples of the Ly$\alpha$ clouds are relatively uniform among
the various samples of high-redshift objects, and Ly$\alpha$ forest lines
are numerous enough to provide statistical analysis, the power spectrum
and higher order statistics of Ly$\alpha$ forests have not been systematically
calculated because of the difficulty in trying to compensate for the
geometry of the samples.

Since FFCs in the range being considered are not strongly affected by the data
outside the range, one can freely extend a sample in spatial range $(D_1, D_2)$
to a larger range $(D_{min},D_{max})$ ($D_{min} < D_1, D_{max} > D_2$) by
adding
zero to the data in ranges $(D_{min}, D_1)$ and $(D_2,D_{max})$. One can
then take statistics by simply dropping all FFCs related to the positions in
the regions of $(D_{min}, D_1)$ and $(D_2,D_{max})$. Using this technique,
all samples can be extended to a desired range $(D_{min},D_{max})$, and the
geometrically complicated samples are regularized.

Local spectrum can also be detected by the Karhuen-Lo\'eve (K-L) transform, or
proper orthogonal decomposition, which is designed for an optimization over
the set of orthogonal transformation of a covariance matrix. However, finding
the K-L eigenvectors of a matrix of order $f$ has computing complexity
$O(f^3)$.
In addition, the K-L bases are not admissible. Even after finding the
eigenvectors of a data set, updating the bases with some extra samples will
cost an additional $O(f^3)$ operations. On the other hand, the DWT can
quasi-diagonalize the covariance matrix, and therefore, K-L transform can
be approximately represented by wavelets which leads to less computing
complexity
(Wickerhauser 1994, Carruthers 1995).

\setcounter{enumi}{5}
\setcounter{equation}{0}

\section{Measures of non-Gaussianity}

\subsection{Spectra of skewness and kurtosis}

For a density (contrast) distribution $\epsilon(x)$ in a range $L=N$ bins,
the non-Gaussianity is usually measured by the skewness $S$ and
kurtosis $K$ defined as
\begin{equation}
S \equiv \frac{1}{N \sigma^3} \sum_{i=1}^{N}
(\epsilon(x_i) - \overline{\epsilon(x_i)})^3,
\end{equation}
and
\begin{equation}
K \equiv \frac{1}{N\sigma^4} \sum_{i=1}^{N}
[(\epsilon(x_i) - \overline{\epsilon(x_i)})^4] - 3.
\end{equation}
These measures cannot describe any possible scale-dependence
of the skewness and kurtosis.

Using the DWT, one can measure the non-Gaussianity by the spectrum of
skewness defined as [see eqs.(3.6)]
\begin{equation}
S_j \equiv \frac{C^3_j}{\sigma^3_j} =
\frac{1}{N_{r} 2^j \sigma^3_j} \sum_{s=1}^{N_r} \sum_{l=0}^{2^j-1}
[(\tilde{\epsilon}_{j,l} - \overline{\tilde{\epsilon}_{j,l}})^3]_s,
\end{equation}
and the spectrum of kurtosis as [eq.(3.7)]
\begin{equation}
K_j \equiv \frac{C^4_j}{\sigma^4_j}
= \frac{1}{N_{r} 2^j \sigma^4_j} \sum_{s=1}^{N_r} \sum_{l=0}^{2^j-1}
[(\tilde{\epsilon}_{j,l} - \overline{\tilde{\epsilon}_{j,l}})^4]_s - 3,
                         \end{equation}
where the variance $\sigma^2_j$ is given by (3.5), and $N_r$ is
the number of samples. Unlike eqs.(5.1) and (5.2), the spectrum description
eq.(5.3) and (5.4)
can detect not only the non-Gaussianity of the density field,
but also the scale of objects contributing to the non-Gaussian behavior.

Note that eqs.(5.3) and (5.4) differ slightly from usual definition of the
skewness or kurtosis by the sum over $s$ from 1 to $N_r$. This is because at
small $j$ one sample on the range $L$ will yield only a small number of
$\tilde{\epsilon}_{j,l}$, makeing the calculation of $K_j$ meaningless.
For instance, for $j = 2$, each sample gives only two FFCs, i.e.
$\tilde{\epsilon}_{2,0} \, \mbox{and} \, \tilde{\epsilon}_{2,1}$. In this
case, $\overline{\tilde{\epsilon}_{j,l}} = (\tilde{\epsilon}_{2,0} +
\tilde{\epsilon}_{2,1})/2 $, and $K_j = -2 $ if $N_r=1$, regardless
whether the sample is Gaussian or what wavelet function is used.
The $K_j$ for each sample can not be calculated separately and still have
meaningful results at lower $j$.
Generally, we have many samples covering the range $L$. One can compile
subsets consisting of $N_r$ samples. The number of $\tilde{\epsilon}_{j,l}$
will then be $N_r$ times larger than one sample making the statistics at
small $j$ viable.  As with the usual definitions of skewness and kurtosis,
$S_j$ and $K_j$ should vanish for a Gaussian distribution.

\subsection{Distribution of clumps}

The effectiveness in detecting the scales of non-Gaussian objects can be
illustrated by clump and valley structures.
Let us consider non-Gaussian density fields consisting of clumps randomly
distributed in a white noise background. Clump distributions are often used
to test methods of detecting non-Gaussianity in large scale structure
study (Perivolaropoulos 1994, Fan \& Bardeen 1995). It has been shown that
one cannot detect the non-Gaussianity of samples by the one-point
probabilities of the individual Fourier modes, even when the samples
contain only a few independent clumps (Kaiser \& Peacock 1991).

To begin, first note that a clump or valley with density perturbation
$\Delta \rho_c$ on length scale $d$ at position $l$ can be described as
\begin{equation}
\rho^{\pm}(x) =
   \left\{ \begin{array}{ll}
      \pm \Delta \rho_c & {\rm if} \ \ lL/2^{J_c} \leq x < (l+1)L/2^{J_c} \\
         0 & {\rm otherwise}
         \end{array}
   \right.
\end{equation}
where $J_c=\log_2 (L/d)$, and the positive and negative signs are for
a clump and a valley, respectively. If a density field $\rho(x)$ consist
of $N$ randomly distributed clumps and valleys of scale $d$, so that
the number density is $N/2^{J_c} d$ on average, the field can be
realized by a random variable $\delta \rho$
with a probability distribution $P[\delta \rho \leq X]$ defined as
\begin{equation}
P[x \leq X]= \left\{ \begin{array}{ll}
                     0           & {\rm if} \ \ X < -\Delta \rho_c \\
                     N/2^{J_c+1} & {\rm if} \ \ -\Delta \rho_c < X < 0 \\
                     1-N/2^{J_c} & {\rm if} \ \ 0 < X < \Delta \rho_c \\
                     1           & {\rm if} \ \ X > \Delta \rho_c
                    \end{array}
           \right.
\end{equation}
The distribution function $\delta \rho$ of clumps and valleys,
$f_c(\delta \rho)$ can then be written approximately as
\begin{equation}
f_c(x)= \frac{dP}{dx}=
(1-\frac{N}{2^{J_c}})\delta(x) +
\frac{N}{2^{J_c+1}}\delta(x -\Delta \rho_c)  +
\frac{N}{2^{J_c+1}}\delta(x +\Delta \rho_c).
\end{equation}
The $\delta(..)$ on the right hand side of eq.(5.7) denote Dirac
$\delta$-functions. The characteristic function of the random variable
$\delta \rho$ of clumps and valleys is
\begin{equation}
\phi_c(u)=\int_{-\infty}^{\infty} f_c(x)e^{i x u} dx
=\frac {2^{J_c}-N}{2^{J_c}} + \frac {N}{2^{J_c}} \cos(\Delta_c u)
\end{equation}
where $\Delta_c=\Delta \rho_c/\bar \rho$. It is very well known that
the ``standard" measures of skewness and kurtosis, i.e. eqs.(5.1) and
(5.2), for the distribution
(5.6) can be calculated from the characteristic function (5.8). The
results are
\begin{equation}
S = - \frac{1}{i\sigma^3}\left [\frac{d^3\phi_c(u)}{du^3} \right ]_{u=0}=0
\end{equation}
and
\begin{equation}
K = \frac{1}{\sigma^4}\left [\frac{d^4\phi_c(u)}{du^4} \right ]_{u=0} - 3
  = \frac{2^{J_c}}{N}-3,
\end{equation}
where
\begin{equation}
\sigma^2= - \frac{d^2\phi_c(u)}{du^2}|_{u=0}
= \frac {N (\Delta_c)^2}{2^{J_c}}
\end{equation}
is the variance of the distribution.

Consider density fields consisting of clumps or valleys randomly
distributed in a background. In this case, the characteristic
function is $\phi(u) = \phi_c(u) \phi_b(u)$, where $\phi_b(u)$ is the
characteristic function of the background distribution. For a
randomly uniform Gaussian background with variance $\sigma_b^2$, the
overall variance is
\begin{equation}
\sigma^2= \frac {N (\Delta_c)^2}{2^{J_c}} + \sigma_b^2,
\end{equation}
and the ``standard" kurtosis is
\begin{equation}
K = \left ( \frac{2^{J_c}}{N} -3 \right )
\left (1+\frac{2^{J_c}}{N (s/n)^2} \right)^{-2},
\end{equation}
where $s/n=\Delta_c /\sigma_b$ is the signal-to-noise ratio. Eq.(5.13)
shows that this distribution becomes Gaussian when $s/n$ is small.

Samples of clumps and valleys randomly distributed in Gaussian noise
background were produced. Figure 5 shows typical fields which
contain A) 16, B) 32, and C) 48 clumps and valleys randomly distributed
in white noise background over the range $L=512$ bins. The signal-to-noise
ratio is $s/n = 2.0$, and the size of the clumps, $d$, is
randomly distributed from 1 to 5 bins, i.e. the average bin width of the
clumps is about 3. Figure 5 shows the spectrum of kurtosis of the these
samples. For comparison, the ``standard" kurtosis, $K$, given by eq.(5.2)
is also plotted at the position $j=9$.

When distributions are non-Gaussian, a Gaussian variance
will no longer be applicable to estimate the statistical errors. Instead,
the error should be calculated from the confidence level of an ensemble of
the samples. In Figure 5, the error bars are the 95\% confidence for
the ensemble consisting of 100 realizations.

Figure 5 shows that the ``standard" measure of kurtosis $K$ is generally
lower than that given by the FFCs, especially when the number of clumps
is large. Moreover, the errors in $K$ are much larger than that of the FFCs,
and $K=0$ is contained in the error bars of $K$. In other words, $K$ is
incapable of detecting the non-Gaussianity of these samples.
This is expected because the measure (5.2) is essentially the same as
that of the MFCs. As mentioned in \S 3, the MFC one-point distributions
will be
Gaussian if the clumps are independent and numerous. On the other hand,
the spectrum of kurtosis confirms the non-Gaussianity of the distribution,
even when the number of clumps is as large as 48. The spectra also show
a peak at $j = 6$ which corresponds to the mean width of the clumps,
$\sim 3$.
\begin{figure}[htb]
\begin{center}
\vspace{-10mm}
\begin{turn}{-90}
\epsfig{file=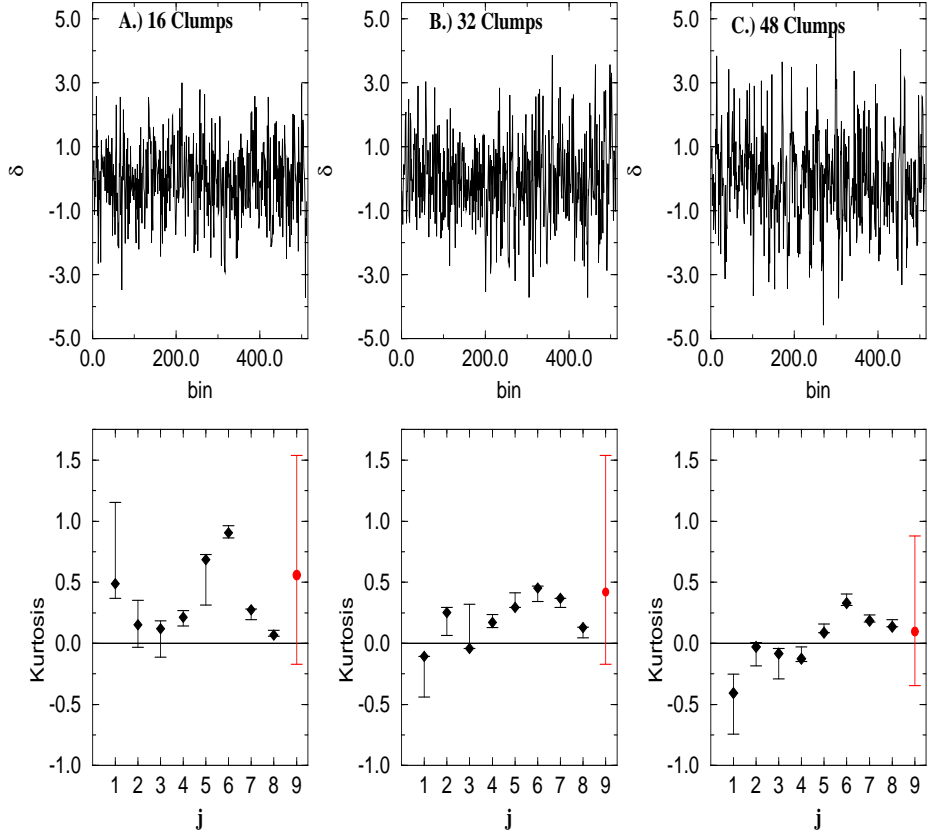,height=4.5in,width=4.5in}
\end{turn}
\end{center}
   \label{fig5}
\vspace{10mm}
\caption{Kurtosis spectra for samples consisting of 16, 32, and
  48 clumps in a length of $L=512$ bins. The sizes of the clumps,
  $d$, are randomly distributed in range of 1 to 5 bins. $S/N$ is 2.}
\end{figure}

\subsection{Suppression of shot noise}

LSS data yield distributions sampled with a finite
number of objects. Such sampling leads to non-Gaussian signals. Even if the
original random field is Gaussian, the sampled data must be non-Gaussian
on scales for which the mean number in one bin is small. This is the
non-Gaussianity of shot noise. Any non-Gaussian behavior of the density
fields will be contaminated by the shot noise.

For instance, in numerical calculations, a distribution $\epsilon(x)$ of
sampled objects is often binned into a histogram, and in order to
maximally pick up information from a real data set, the bin size is taken
to be the resolution of the coordinate $x$. However when the bin size of
the histogram is less than
the mean distance of neighbor objects, the value of the binned
$\epsilon(x)$ will typically be 0 or 1. Thus, the sample is actually a
d=1 clump distribution with a one point distribution given
by eqs.(5.5) and (5.6). It is {\em not} a Gaussian distribution.
Maximally picking up information about the distribution comes at a cost of
artificially introducing non-Gaussian behavior.

It is difficult to separate the non-Gaussianities caused by shot
noise and binning with that given by non-Gaussian structures. The measure
$K$ contains all contributions to the non-Gaussianity of the density field.
As a consequence, $K$  will be sampling-dependent, and not suitable
for confronting observation with theory, or for discriminating among models.

However,  non-Gaussian shot noise or binning have different
spectrum features from that of non-Gaussian samples. For shot noise, the
distributions on large scales is a superposition of the small scale field.
According to the central limit theorem the non-Gaussian shot noise
will rapidly approach zero on larger scales. In other words, if the mean number
in a bin is larger for larger bins, the one-point distribution of shot
noise will become Gaussian on larger scales. The values of $|K_j|$
for shot noise should rapidly approach zero as $j$ gets smaller,
i.e. their kurtosis spectrum should be monotonously decreasing with
decreasing $j$.

To illustrate this point, Figure 6 plots the spectrum of kurtosis for a
random white noise sample given by
\begin{equation}
x_i = x_1 + (x_2 - x_1)\cdot RAN
\end{equation}
where $x_i$ is the position of i-th object, and $RAN$ is random
number in (0, 1). We take the size of the sample $(x_2 - x_1) = 64$
bins, and the total number of objects is 120. Figure 6 shows that the
non-Gaussianity of the random sample is significant only
when the mean number of objects per bin is less than 2.

The non-Gaussianity of the shot noise will significantly
be suppressed in the spectrum of FFC cumulants with increasing scale.
This feature is useful in distinguishing between non-Gaussian
clumps and shot noise. For instance, one can definitely conclude the
existence of non-Gaussian structures if the kurtosis spectrum
is flat, or contains peaks as in Figure 5.
\begin{figure}
\begin{center}
\begin{turn}{-90}
\epsfig{file=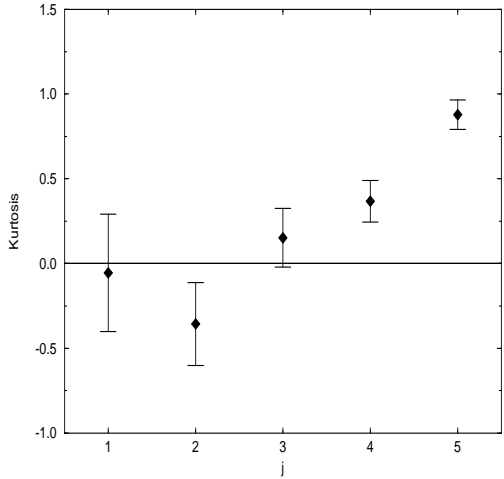,height=3in,width=3in}
\end{turn}
   \label{fig6}
\caption{Kurtosis spectrum of 120 objects randomly
distributed in 64 bins.}
\end{center}
\end{figure}

\setcounter{enumi}{6}
\setcounter{equation}{0}

\section{A preliminary result using DWT: Ly$\alpha$ forests}

\subsection{LSS Problems of Ly$\alpha$ clouds}

Ly$\alpha$ absorption line forests in QSO spectra come from intervening
absorbers, or clouds, with neutral hydrogen column densities ranging from
about $10^{13}$ to $10^{17}$ cm$^{-2}$ at high redshifts. Since
the size of the Ly$\alpha$ clouds at high redshift is as large as
100 - 200 h$^{-1}$ Kpc, and their velocity dispersion is as low as
$\sim$ 100 km s$^{-1}$ (Bechtold et al. 1994, Dinshaw et al. 1995,
Fang et al. 1996), it is generally believed that
the Ly$\alpha$ clouds are probably neither virialized nor completely
gravity-confined, but given by pre-collapsed areas in the density field.
It is reasonable to assume that the Ly$\alpha$ clouds joined the
clustering process in the universe.

However, almost all of the results drawn from two-point correlation function
analysis of Ly$\alpha$ forest lines have failed to detect any significant
clustering on the scales less than 10 h$^{-1}$ Mpc (Weymann 1993), where
$h$ is the Hubble constant in the unit of 100 km s$^{-1}$ Mpc$^{-1}$.
In other words, the power spectrum of the Ly$\alpha$ clouds is flat, i.e.
 similar to white noise.

Theoretically, it is hard to  believe that the distribution of the Ly$\alpha$
lines is only white noise since the Ly$\alpha$ clouds should be formed via
the same process as that for other objects, i.e., gravitational clustering.
In fact, contrary to the results of two point correlation function,
statistics based on other methods show definite deviations
of the Ly$\alpha$ forests from a uniform random distribution.
For instance, the distribution of nearest neighbor
Ly$\alpha$ line intervals was found to be significantly different from a
Poisson distribution (Liu and Jones 1990). Based on the Kolmogorov-Smirnoff
(K-S) statistic, Ly$\alpha$ absorbers have been shown to deviate from a uniform
distribution at $\sim 3\sigma$ significant level (Fang, 1991).

It is of fundamental importance to understand why the two point correlation
function, which is one of the most used ways to detect structure, fails to
detect any clustering when other methods are showing definite structure.
Two possible explanations are 1.) the clustering cannot be detected by
the two-point correlation function on scale less than 10 h$^{-1}$ Mpc,
2.) the clustering
cannot be detected by second order statistics. We look at these in more
detail.

1. If the spectrum of Ly$\alpha$ clouds is different from white noise only
on large scales, the clustering will be missed by the
two point correlation function.
As we know, the mean number density of the Ly$\alpha$ clouds significantly
evolves with redshift. The evolution is generally described as
\begin{equation}
\frac{dN}{dz}=\left (\frac{dN}{dz} \right )_0(1+z)^{\gamma}
\end{equation}
where $(dN/dz)_{0}$ is the number density extrapolated to zero redshift,
and $\gamma \sim 2$ the index of evolution. No mean density is
available for calculating the two-point correlation function.
Since the correlation amplitudes are less than the uncertainty
of mean density of Ly$\alpha$ lines on scales larger than 5 h$^{-1}$ Mpc,
the two-point correlation will overlook structures on large scales.

2. If the clustering of Ly$\alpha$ clouds cannot be detected by
second order statistics, the two-point correlation function will certainly
be incapable of detecting structures.
Using a linear simulation of density fields, it is found that the
simulated clouds indeed show no power of their two-point
correlation functions on scales from about 100 km s$^{-1}$ to 2000 km
s$^{-1}$ (see Figure 7) (Bi, Ge \& Fang 1995, hereafter BGF). Since the
perturbation spectrum used for the simulation is not white noise,
the distribution of the clouds should contain large scale structures.
No power in the line-line correlations may not mean that
the distribution is white noise, but instead indicate, that the two-point
correlation
function sometimes is ineffective in detecting structures on large
scales.
\begin{figure}
\begin{center}
\epsfig{file=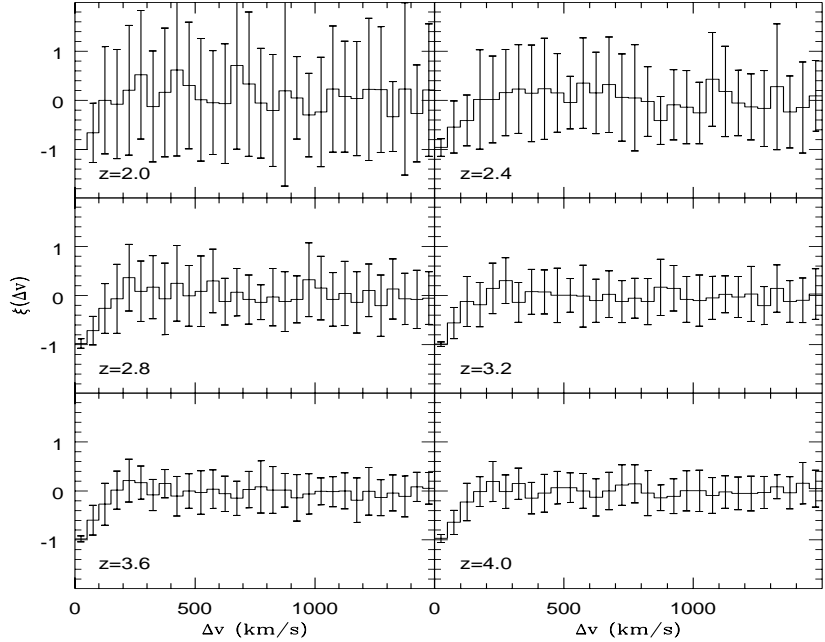,height=4.5in,width=4.5in}
\vspace{-1in}
   \label{fig7}
\caption{Two-point correlation functions of the BGF simulated
sample of Ly$\alpha$ forests with $W \geq 0.16 \AA$ in the LCDM
. The error bars represent 1 $\sigma$.}
\end{center}
\end{figure}

In order to clarify  clustering of the Ly$\alpha$ clouds we need to
determine whether: A.) does the power spectrum of Ly$\alpha$ clouds stay
flat on large scales? B.) is the clustering of the clouds detectable by
higher order statistics?

\subsection{Flatness of power spectrum}

Question A can be answered by a DWT spectrum detection. We will look at
two popular real data sets of the Ly$\alpha$ forests. The first was
compiled by  Lu, Wolfe and Turnshek (1991, hereafter LWT). The total
sample contains $\sim$ 950 lines from the spectra of 38 QSO that exhibit
neither broad absorption lines nor metal line systems. The second set is
from Bechtold (1994, hereafter JB), which contains a total $\sim$ 2800 lines
from 78 QSO's spectra, in which 34 high redshift QSOs were observed at
moderate resolution. To eliminate the proximity effect, all lines with
$z \geq z_{em} - 0.15$ were deleted from our samples. These samples
cover a redshift range of 1.7 to 4.1, and a comoving distance range from
about $D_{min}$=2,300 $h^{-1}$Mpc to $D_{max}=$3,300 $h^{-1}$Mpc, if
$q_{0} = 1/2$.

For comparison, we also work on the simulation samples of  BGF. The density
fields
in this simulation are generated as Gaussian perturbations with a linear
power spectrum given by the standard cold dark matter model (SCDM), the cold
plus hot dark  matter model (CHDM), and the low-density flat cold dark matter
model (LCDM). The baryonic matter is assumed to trace the dark matter
distribution on scales larger than the Jeans length of the baryonic gas,
but is smooth over structures on scales less than the Jeans length. Within a
reasonable range of the UV background radiation at high redshift, the
absorption of the pre-collapsed baryonic clouds is found to be in good
agreement with the features of Ly$\alpha$ forest. In particular, the
LCDM gives good fits to: 1) the number density of Ly$\alpha$
lines; 2) the distribution of equivalent width; and 3) the
redshift-dependence of the line number.

Because different QSO's forest of real samples cover different
redshift ranges, it is not trivial to directly find the Fourier power spectrum
from an ensemble consisting of such geometrically complex forests. However,
this complex geometry can easily be regularized by the method described in \S
3.
Using this technique, all samples were extended in comoving space, to cover
1024 bins with each bin of comoving size $\sim 2.5$ h$^{-1}$ Mpc. Thus, all
QSO samples were treated uniformly. The
Fourier spectrum can then be detected by the FFCs. When we are only
interested in the shape, not the amplitude of the spectrum, the result
is independent of the uncertainty of overall mean density.

\begin{figure}
\begin{center}
\vspace{-3.25cm}
\epsfig{file=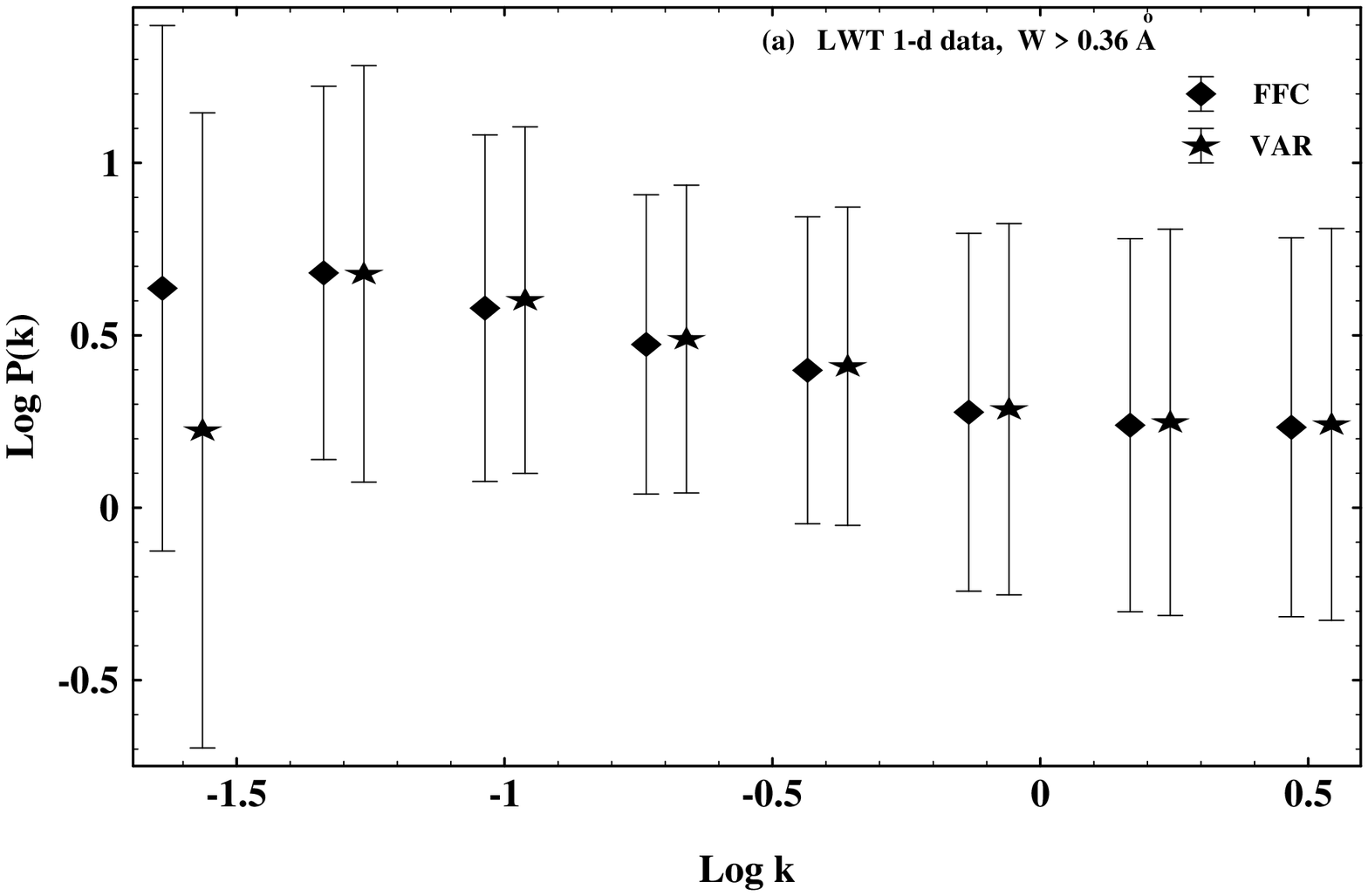,height=3.50in,width=4in}

\vspace{-3.25cm}
\epsfig{file=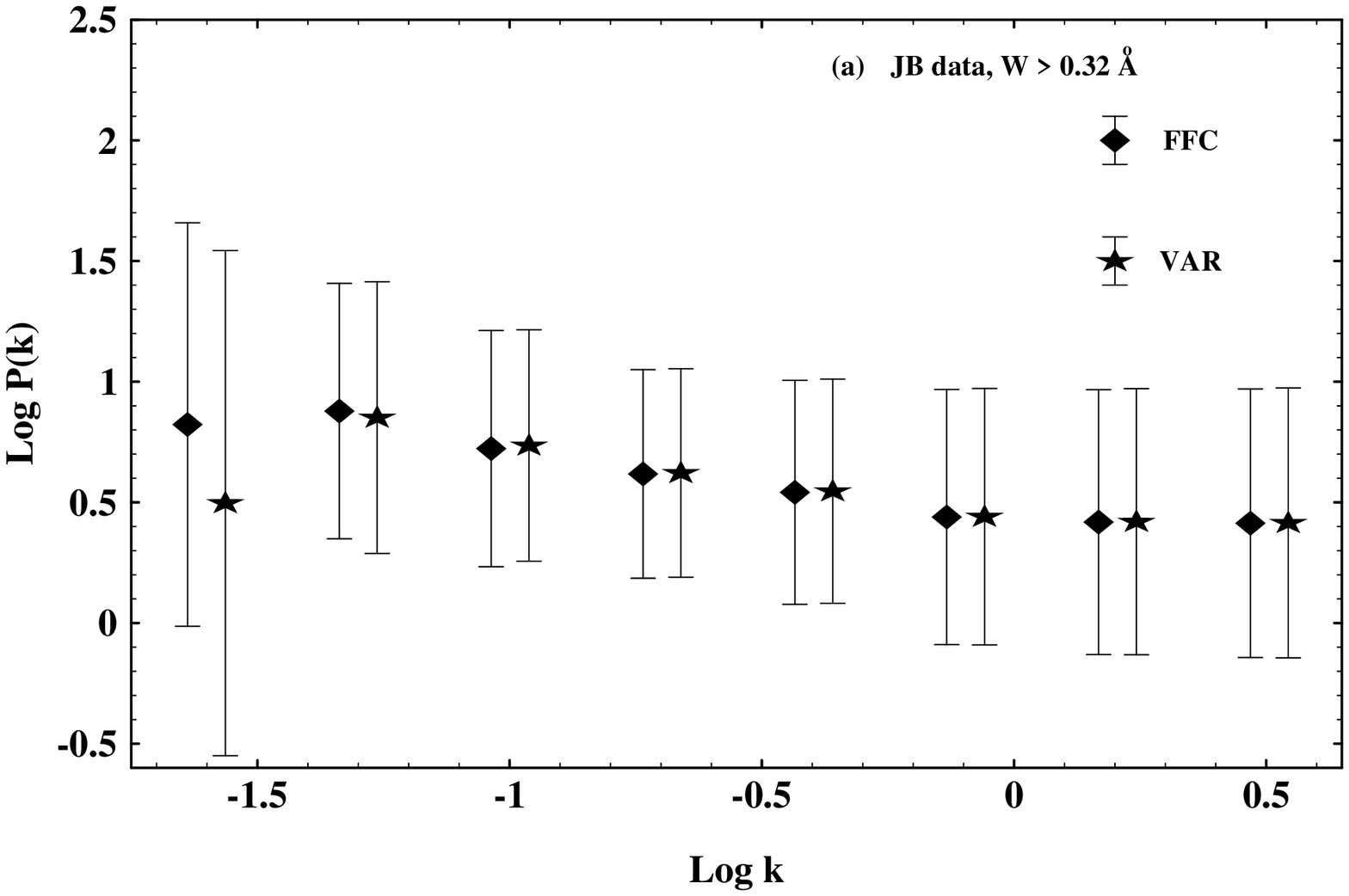,height=3.50in,width=4in}

\vspace{-3.25cm}
\epsfig{file=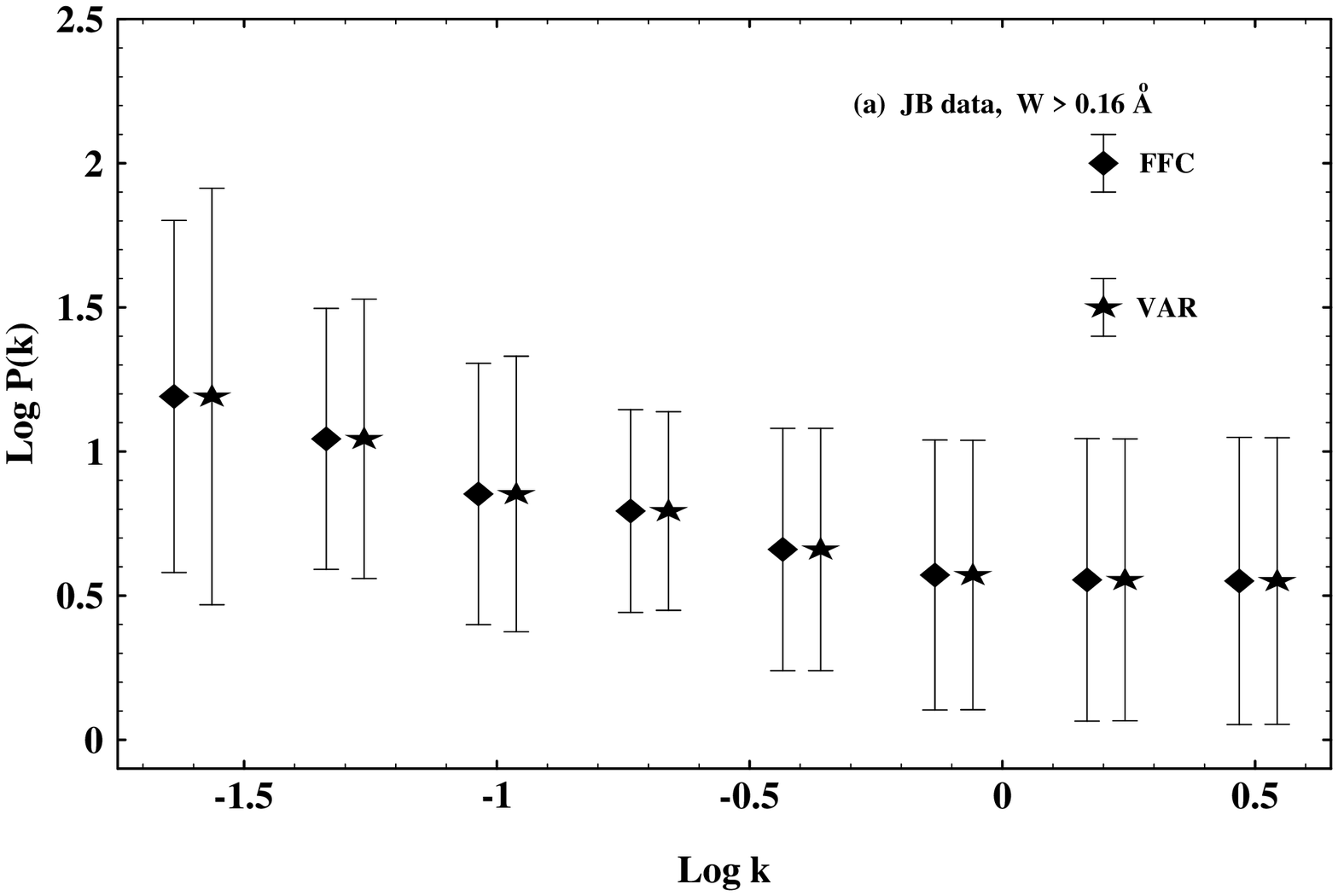,height=3.50in,width=4in}

\vspace{-2.25cm}
\end{center}
\label{fig8}
\caption{1-D spectra $P(k)_j$ (diamond) and $P^{var}(k)_j$
(star) of a.) LWT Ly$\alpha$ forest samples with width $>$ 0.36 $\AA$;
b.) JB with  $W >0.32 \AA$ and c.) JB with  $W> 0.16 \AA$. For clarity,
the points $P^{var}(k)_j$ are plotted at $\log k + 0.05$.}
\vspace{-1mm}
\end{figure}

The spectra of LWT($W>0.36\AA$), JB($W>0.32\AA$) and JB($W>0.16\AA$)
in the entire redshift range $1.7<z<4.1$ are plotted in Figures 8a, b, c,
respectively. For comparison, Figure 9 shows the spectra of LWT($W>0.36 \AA$),
JB($W>0.32 \AA$), SCDM and CHDM with  $W>0.32 \AA$. The error bars in the
Figures 8, 9 come from the average over the samples of QSO's absorption
spectrum. The errors at large scale are about the same as that on small
scales. This means that the spectrum can uniformly be detected on scales as
large as about 80 h$^{-1}$ Mpc by the FFCs.

All spectra in Figures 8 and 9 are rather flat on the range of $\log k > -1$,
i.e. on scales less than 5 h$^{-1}$ Mpc, and slightly increases with scale in a
range of 10 - 80 h$^{-1}$ Mpc ($\log k \leq -1$). These results are
consistent with the result of no correlation power on scales less than
10 h$^{-1}$ Mpc. The flatness of these spectra can be described by the power
law index $\alpha$, which is found by fitting the observed spectra with
power law $P(k) \propto k^{\alpha}$. The results are:
$\alpha=-0.26 \pm 0.42$ for the LWT ($W>0.36 \AA$),
$\alpha= -0.23 \pm 0.41$ for the JB ($W>0.32 \AA$), and
$\alpha= -0.23 \pm 0.37$ for the JB $(W>0.16 \AA)$. The two independent data
sets, LWT and JB, show almost the same values of the index $\alpha$. This
strongly implies that this feature is common among the Ly$\alpha$ forests.

One can conclude that although the power spectrum seems to increase
on large scales, as discussed in \S 6.1, the values of
$\alpha$ and its errors show that the distribution of the
real sample is consistent with a flat spectrum ($\alpha \sim 0$) on scales
less than 100 h$^{-1}$ Mpc. Moreover, the power spectra of the simulated
samples
of the SCDM, CHDM and LCDM also are quite flat, i.e.
$\alpha = -0.93 \pm 0.15$ for SCDM ($W>0.16 \AA$) and
$\alpha = -1.06 \pm 0.08$ for CHDM ($W>0.16 \AA$).  It appears that the
difference in the results drawn from the 2-point correlation function and other
statistical methods is not due to the possibility that the structures occur on
large scales.
\begin{figure}
\begin{center}
\vspace{-0.1in}
\epsfig{file=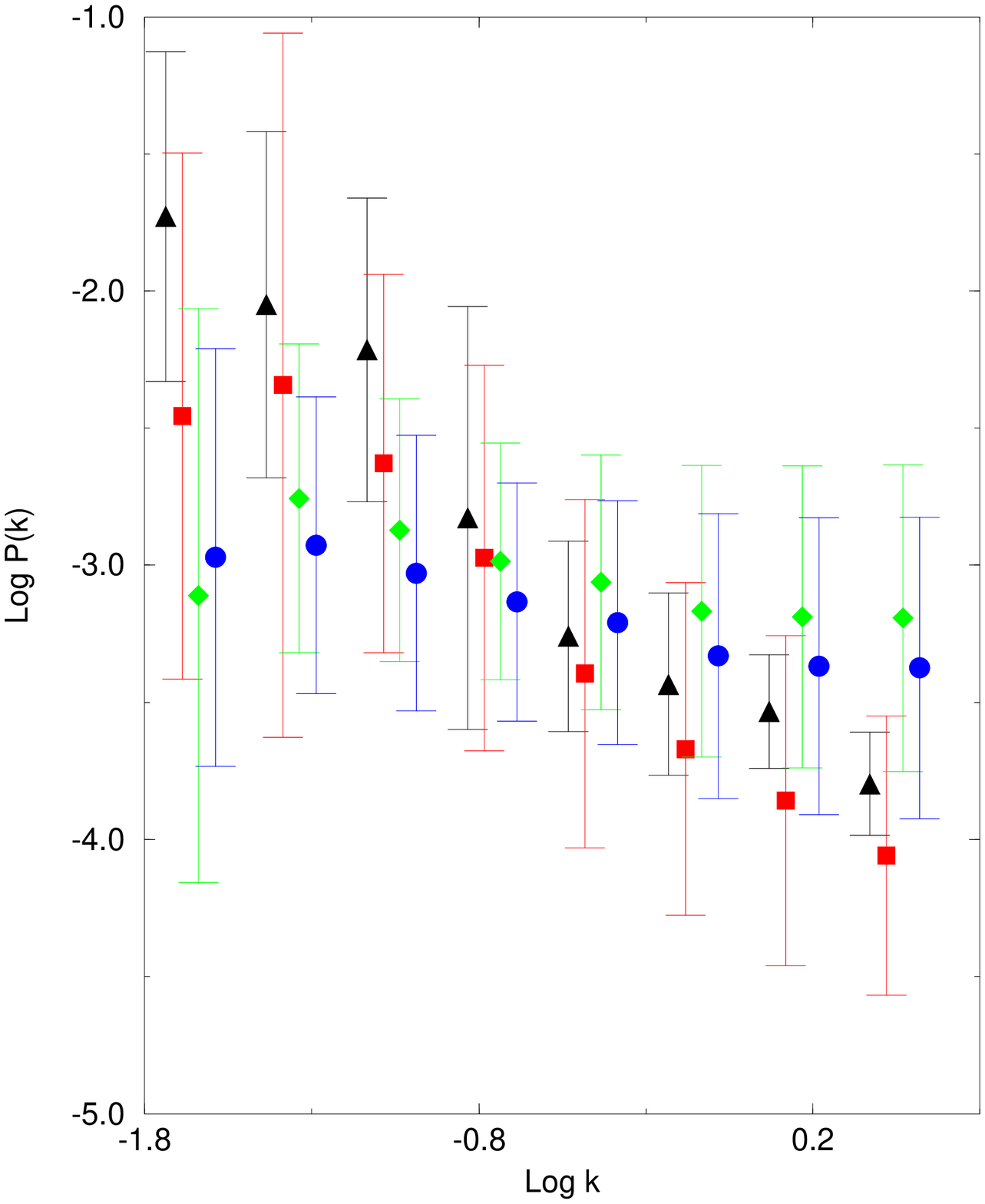,height=4in,width=4in}
\end{center}
   \label{fig9}
\vspace{-15mm}
\caption{1-D spectra $P(k)_j$ of LWT Ly$\alpha$ forest samples
with width $>$ 0.36 $\AA$ (circle); JB samples with $W>$ 0.32 $\AA$ (diamond);
SCDM with $W>0.32 \AA$ (triangle); CHDM with $W>0.32 \AA$ (square).
The spectrum is given by data in the entire redshift range $1.7<z<4.1$.
$k$ is in unit h Mpc$^{-1}$. The error bars are obtained from the average
over the samples of QSO's absorption spectrum. For clarity, the four
types of points are slightly shifted from each others along the $k$ axis.}
\end{figure}

\subsection{Kurtosis on large scales}

 From the flatness of the spectrum, it should not be concluded that the
distribution of Ly$\alpha$ clouds must be white noise. Instead, this may
indicate that the power spectrum and two-point correlation function are not
suitable for describing the statistical features of the system being
considered. Statistically, it is essential to measure non-Gaussianity in
order to detect the clustering of density field with flat power spectrum.

This point can be illustrated by the simulation sample BGF. As shown in
Figure 7, the power spectrum of the SCDM  (W$>0.36 \AA$) is flat.
The spectrum is almost the same as that of a random sample
generated by eq.(5.14) with the  redshift-dependence of the
number density of the Ly$\alpha$ lines included.  That is, in each
redshift range $\Delta z = 0.4$ the number of lines in the random
sample is taken to be the same as the SCDM sample.

\begin{figure}
\begin{center}
\vspace{-15mm}
\begin{turn}{-90}
\epsfig{file=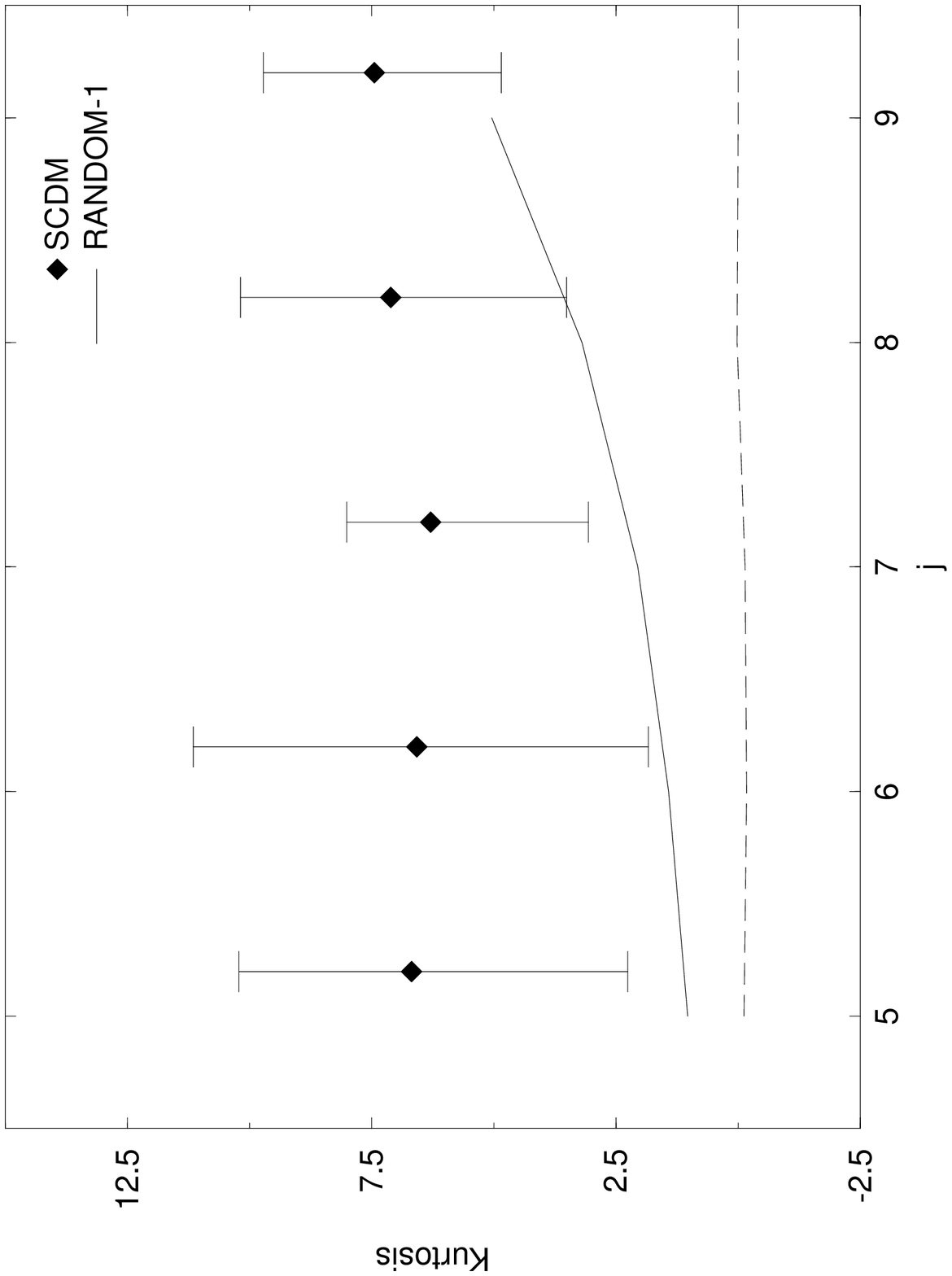,height=3in,width=3in}
\end{turn}
   \label{fig10}
\caption{Kurtosis spectra of BGF sample of SCDM model
and random data (solid line), respectively. Dot line is for $K_j=0$.}
\end{center}
\end{figure}
However, Figure 10 shows that the amplitude of the kurtosis spectrum for the
SCDM sample is systematically larger than the random sample. Recall that the
error bars in Figure 10 do not represent the 1 $\sigma$ Gaussian errors, but
the 95\% confidence level from the ensemble of the samples. The difference
between the spectra of the SCDM and random sample is significant.
The kurtosis of the random sample is completely given by the shot noise.
As expected, the kurtosis of the random sample is monotonously decreasing
with decreasing of $j$.

In trying to apply this technique to observational data, one immediately
encounters a serious problem.
In order to compute the skewness and kurtosis spectrum of real data,
subsets of the samples are needed for eqs.(5.3) and
(5.4). Unlike the simulated
samples, where as much data as needed can be generated, the available real
data are limited. There are only $N_r=$ 43 samples (forests) for LWT and 78
for JB. In order to
effectively use this data, $M \leq N_r$ files from among the complete $N_T$
samples are chosen to form a subset.  Various combinations of the subsets $M$
are then combined to form an ensemble. To investigate the effect of different
combinations, the subsets $M$ are formed by varying the total number of files
chosen from the parent distribution, $N_r$, as well as changing the order in
which the individual files are selected. It is found that the skewness and
kurtosis calculated from these $M$-file ensembles are very stable until $M$
contains as few as 7 or 8 files, i.e. until only approximately 5\% of the
total lines remain in the subset. The 95 \% confidence intervals are then
estimated from the ensembles.
\begin{figure}
\begin{center}
\vspace{-15mm}
\begin{turn}{-90}
\epsfig{file=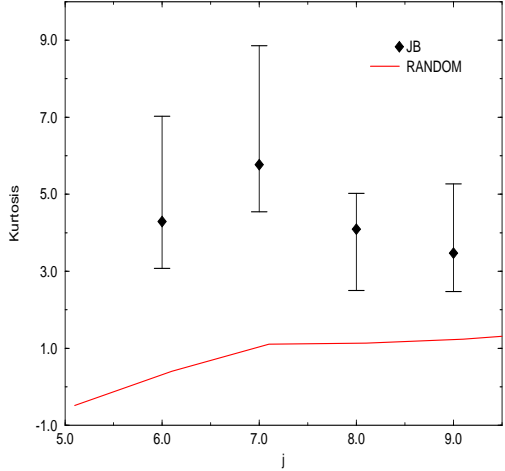,height=3in,width=3in}
\end{turn}
\end{center}
   \label{fig11}
\caption{Kurtosis spectra for JB data (W$>0.16\AA$),
and random sample (dashed line).}
\end{figure}

Figure 11 shows the kurtosis spectra of the JB data and its random sample.
The difference between JB sample ($W>0.16 \AA$) and random sample is more
significant than the difference between SCDM and random data.
The amplitudes of the kurtosis spectrum are much higher than that of random
data on scales larger than 10 h$^{-1}$Mpc. Figures 12
and 13 show the skewness and kurtosis spectra of the LWT ($W>0.36 \AA$) and
JB ($W > 0.32 \AA$) data sets. Again, the two independent data sets show the
same
statistical features. One can conclude that the clustering
of the Ly$\alpha$ clouds can only be clearly described by higher (than 2) order
statistics. The distribution of the clouds is non-Gaussian on all scales
being detected, i.e. less than 80 h$^{-1}$Mpc.
\begin{figure}
\begin{center}
\vspace{-35mm}
\begin{turn}{-90}
\epsfig{file=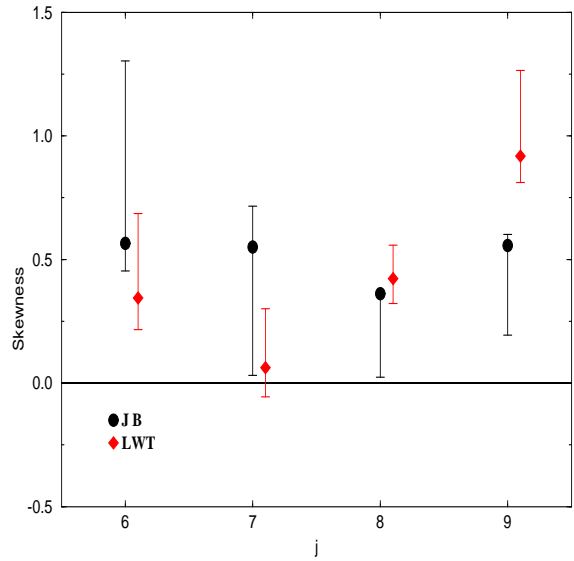,height=3.5in,width=3.5in}
\end{turn}
\end{center}
   \label{fig12}
\caption{Skewness spectra for samples of LWT (W$>0.36\AA$) and
JB (W$>0.32\AA$).}
\end{figure}
\begin{figure}
\begin{center}
\vspace{-1in}
\vspace{-30mm}
\begin{turn}{-90}
\epsfig{file=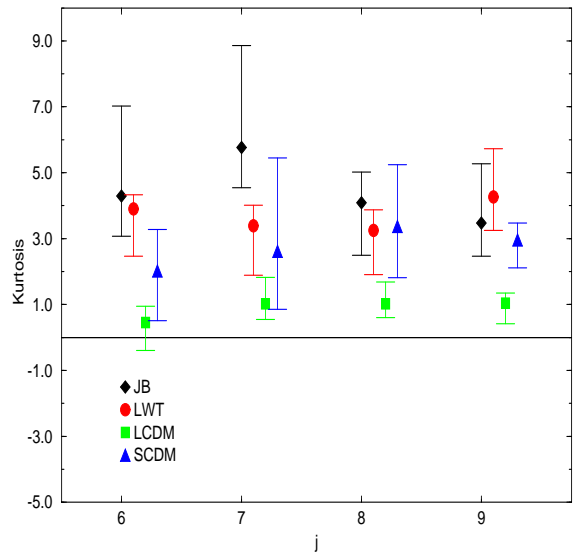,height=3.5in,width=3.5in}
\end{turn}
\end{center}
   \label{fig13}
\caption{Kurtosis spectra for samples of LWT (W$>0.36\AA$) and
JB (W$>0.32\AA$).}
\end{figure}


\subsection{Structures identified by multiresolution analysis}

Direct identification of structures is important in the description of
LSS, because it allows us to ``see" the clustering of object's
distribution and to compare the clusters of simulated samples with
observations.

For weakly clustered distributions, it is difficult, even impossible, to
distinguish the clusters caused by dynamical interaction with that of random
fluctuations. Many clusters are identified at the 2 or 3 $\sigma$ confidence
levels. If the identification is done on a large number of realizations,
some $3 \sigma$ events may be due to random processes.
We will discuss a method of identifying structures by a DWT
multiresolution, in particular, how to handle the random fluctuations in
order to get robust conclusions even when the identified clusters are
contaminated by random fluctuations.

\begin{figure}
\begin{center}
\epsfig{file=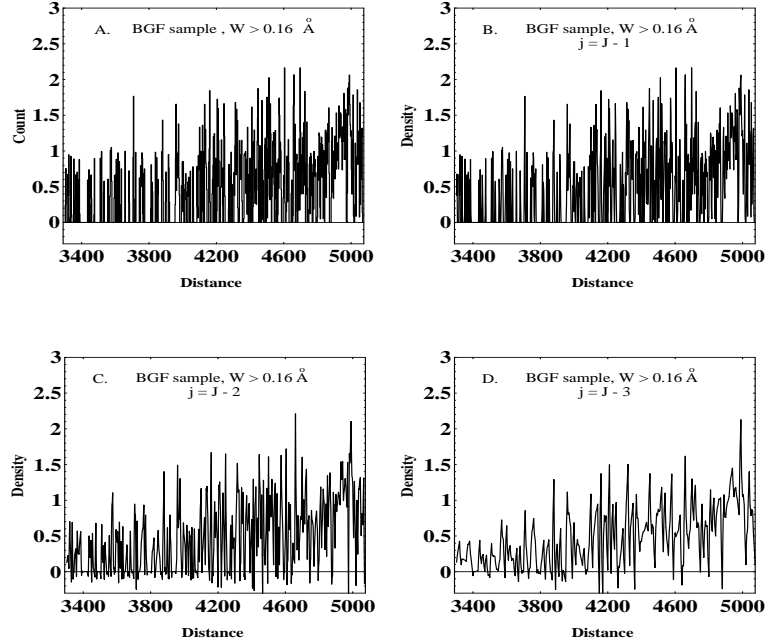,height=4in,width=4in}
\end{center}
   \label{fig14}
\vspace{-15mm}
\caption{A section of the wavelet reconstruction of density fields
for the BGF sample with $W>0.16 \AA$. A. the original (or scale $j=10$)
line distribution.  B. C. and D. the reconstructed fields for scale $j=9$,
8, and 7, respectively.  The distance is in units of h$^{-1}$Mpc.}
\end{figure}
The DWT multiresolution is based on eqs.(2.20) and (2.48). For a given
distribution $\epsilon(x)$, one can decompose it into $\epsilon^j(x)$ on
various scales. As an example, Figure 14 shows the original distribution of
the LCDM ($W>0.16\AA)$ ($j=10$), and the result of the reconstruction on
$j=9, 8$  and 7, corresponding to scales (in comoving space) of about 5, 10
and 20 h$^{-1}$ Mpc, respectively.

The peaks in the field $\epsilon^j(x)$ correspond to high
density regions, and are possible clusters on  scale $j$.
Various methods of structure identification are designed to pick out such
high density regions. Since the MFCs, $\epsilon_{j,l}$, represent the
density of the field at position $l$ and on scale $j$, we can directly
identify possible clusters by picking out the peaks of the MFC distribution.
Figure 15 shows a section of the MFC reconstructed distribution for the
forest of QSO-0237. The error bars in Figure 15 are 1-$\sigma$ calculated
from 100 uniformly random samples which match the number of lines in
each redshift interval, $\Delta z = 0.4$, of the parent sample QSO-0237.
The peaks of $\epsilon_{j,l}$ distribution are identified as clusters of
Ly$\alpha$ clouds on scale $j$ and position ($l$). The strength or richness
of the clusters can be measured by $R=\epsilon_{j,l}/\sigma$.

\begin{figure}
\vspace{-.25in}
\begin{center}
\epsfig{file=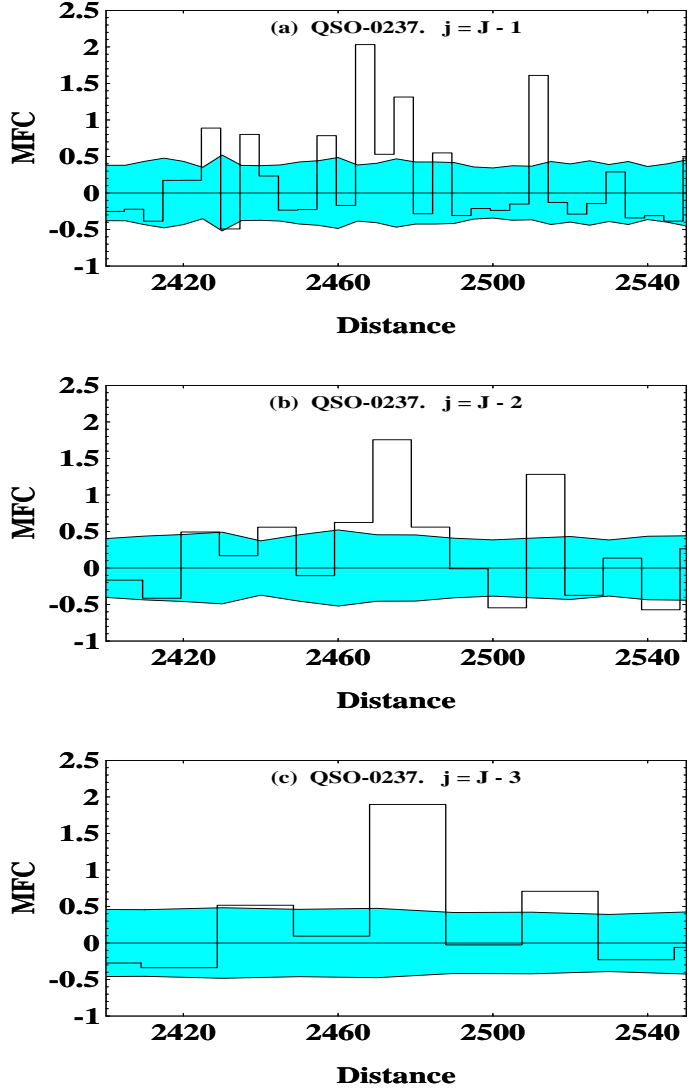,height=6in,width=6in}
\end{center}
   \label{fig15}
\caption{A section of the MFC coefficients
of QSO-0237 Ly$\alpha$ forest with line width $W>0.16 \AA$a BGF and
the corresponding random samples. A. B. and C. are for scales $j=9$, 8 and
7, respectively. The
distance is in units of $h^{-1}$Mpc. The error bars represent one $\sigma$
given by 100 random samples.
It is interesting to note that the $j=9$ clusters shown around
2465-2480 h$^{-1}$ Mpc also appear as $j=8$, 7  clusters at the same place.
That is, this structure appears at all three resolution scales. On the
other hand, the structure appearing at 2505-2520 Mpc only appears on the
scales $j= 9$ and 8, but not on larger scales.
}
\end{figure}

Using this identification scheme, one can count the number strength
distribution of
clusters, $N_j(>R)$, which is defined as the total number of the clusters
on scale $j$ with strength larger than a given $R$. For the BGF sample of
the LCDM ($W < 0.16 \AA$), the $N_j(>R)$  on scales $j=9,$ 8 and 7 is
plotted in Figure 16, in which the error bars are given by the average
among the 20 BGF simulated samples. For the JB ($W>0.16 \AA$), the results
of $N_j(>R)$ are shown in Figure 17.

The advantage of using the DWT to identify clusters is the ability to
systematically study
clusters on all scales and with various strengths. It is easy to estimate
the number of clusters which are due to fluctuations, because the
mother functions
$\phi_{j,l}$ are orthogonal with respect to $l$ and each position
$l$ corresponds to an independent realization. For instance, when $j=9$,
the MFCs
$\epsilon_{j,l}$ are detected from $2^9=1024$ realizations. In this case
the number of $R \leq 2$ events for white noise should be about
$1024 \times 0.03 \sim 30$. The shape of the function $N_j(>R)$ is more
important. In the case of white noise, the number strength
function normalized at $R_0$, i.e. $N_j(>R)/N_j(>R_0)$, should be
${\rm erfc}(R)/{\rm erfc}(R_0)$, ${\rm erfc}(x)$ being the
complementary error function. Again, for white noise, the number-strength
distributions on different scales should be satisfied the relation
$N_j(>R) \sim 2^nN_{j-n}(>R)$.

 From Figures 16 and 17, it is easy to see that for both the BGF and JB data,
we have $N_j(>R)/N_j(>2) > {\rm erfc}(R)/{\rm erfc}(2)$, and
$N_j(>R) > 2^nN_{j-n}(>2)$ for all $R>2$. Both samples
contain clusters not originating from random fluctuations.

\vspace{.40in}

\begin{figure}[htb]
\vspace{2cm}
\centering\epsfig{file=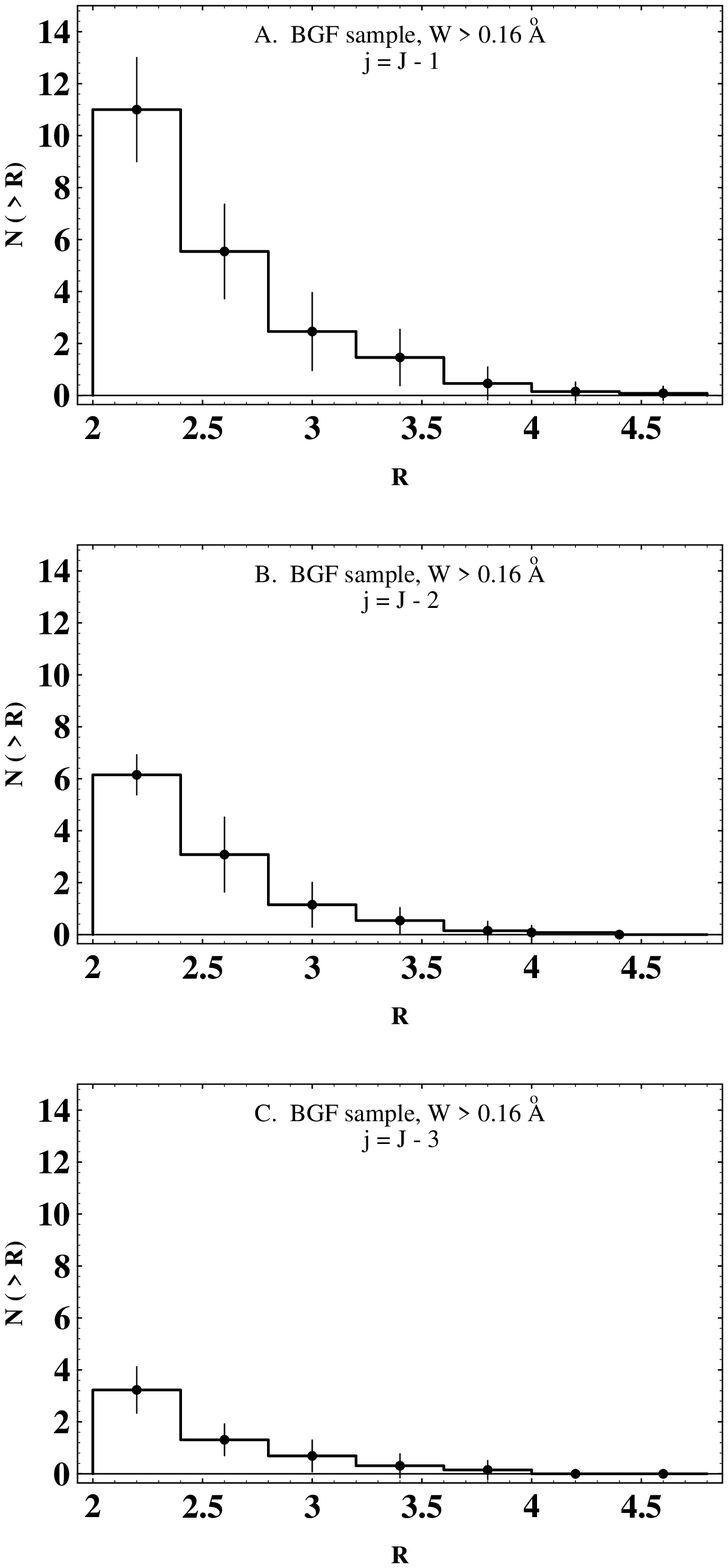,height=4in,width=4in}
   \label{fig16}
\caption{The number of clusters $N_j(>R)$ identified
from BGF samples, where $R$ is the richness of the clusters in units
of $\sigma$. A. B. and C. are for scales $j=9$, 8 and 7,
respectively. The error comes from average among 20 BGF samples.}
\end{figure}
\begin{figure}[htb]
\vspace{2cm}
\centering\epsfig{file=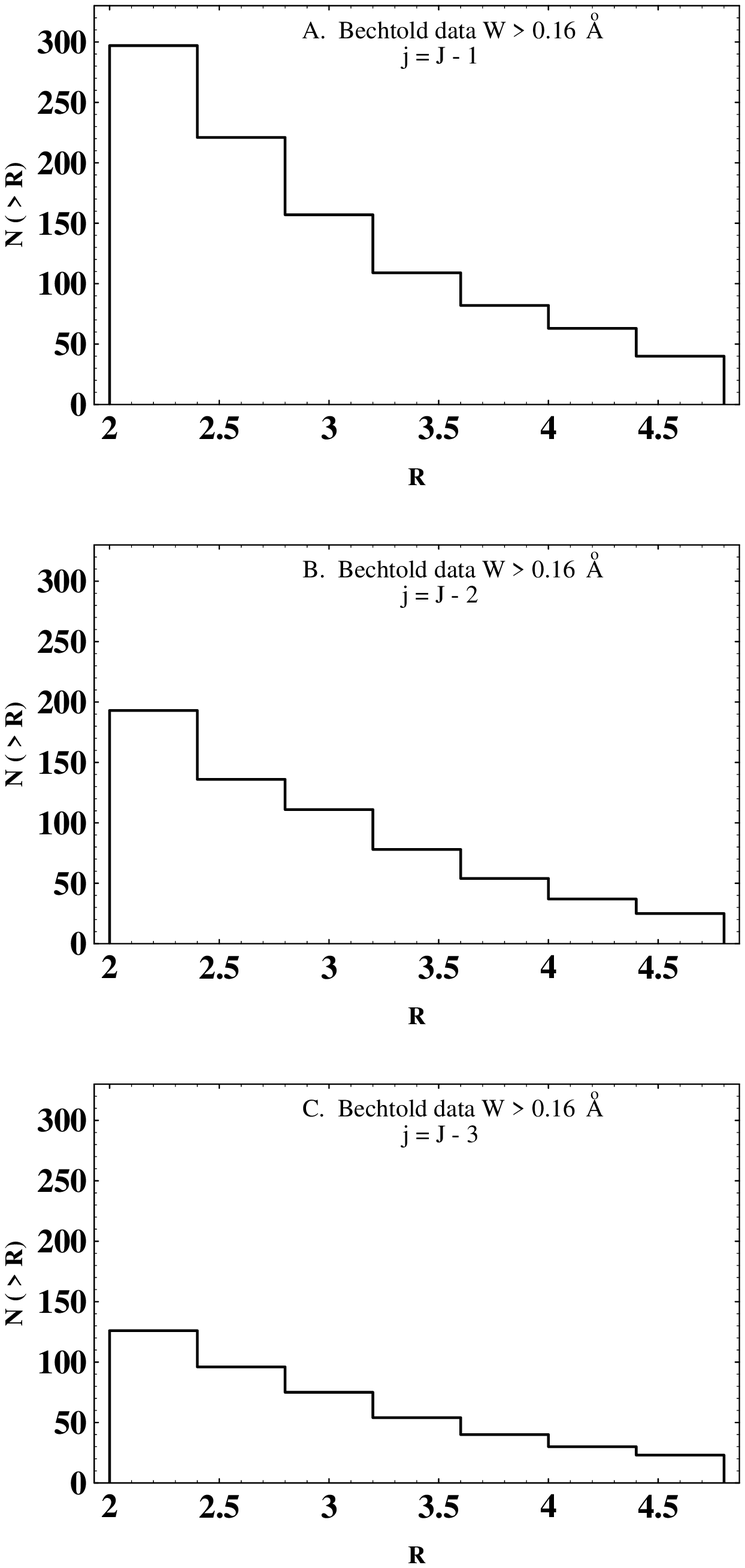,height=4in,width=4in}
   \label{fig17}
\caption{Number of clusters $N_j(>R)$ identified
from the JB $W>0.16\AA$, where $R$ is the richness of the clusters
in units of $\sigma$. A. B. and C. are for scales $j=9$, 8 and 7,
respectively.}
\end{figure}

\subsection{Discrimination among models}

Since the DWT can uniformly calculate statistical quantities on different
scale and different orders, it provides a more powerful tool to dissect models.
In the last few sections, we have performed this dissection on the BGF samples.
As mentioned in \S 6.1 these samples are linear. Nonetheless using
traditional tests, such as the number density, two-point correlation function,
the Gunn-Peterson effect etc, they show the same behavior as the observational
data.  However, these tests involve at most, second
order statistics.  If the Ly$\alpha$ clouds have undergone
non-linear clustering, we should expect that the BGF samples should not compare
with the real data when compared using higher order tests. The results of
\S 6.2 and 6.3, do indeed, confirm this.  That is,
in terms of second order statistics, the BGF data behaves like the real data,
but for higher order tests, the BGF shows significant differences from the LWT
and JB data sets.

For instance, the ratios of $N_j(>R)/N_{j'}(>R)$ and
$N_j(>R)/N_{j}(>R')$ are very dependent
on non-Gaussian (i.e. higher order) behavior (\S 6.3) and so it can also be
used
as a discriminator. From Figures 16 and 17, one can find that
$N_{9}(>3.5)/N_9(>2) =  (7 \pm 1.5)\%$ for BGF sample, but
$N_{9}(>3.5)/N_9(>2) = 23\%$ and $28\%$ for the LWT and JB data, respectively.
The difference of $N_j(>4)/N_j(>2)$ between real and simulated sample is
more remarkable. Almost no $ R > 4.5$ clusters are detected in BGF
samples, while they exist in the real sample.

We can recast these results by looking at the redshift-dependence of the number
of clusters.
As in eq.(6.1), we analyzed the evolution of the number density of clusters
on scales $j=9,$ 8, and 7. The results are plotted in Figure 18. Figure 18a
is for LWT ($W > 0.36 \AA$), and 18b for JB ($W > 0.32\AA$). The two data
sets showed, once again, the same features. The top curves in Figures 18a
and b are the original results (number density of Ly$\alpha$ lines) of LWT
and JB. This number density increases with redshift. However, the number
densities of $j=9$, 8 and 7 clusters show an opposite evolution, decreasing
with increasing redshift. That is, large scale structures traced by
Ly$\alpha$ lines were growing from the era $z=4$ to $2$. This evolutionary
feature was never revealed before the DWT analysis.
\begin{figure}[hbt]
\vspace{5mm}
\epsfig{file=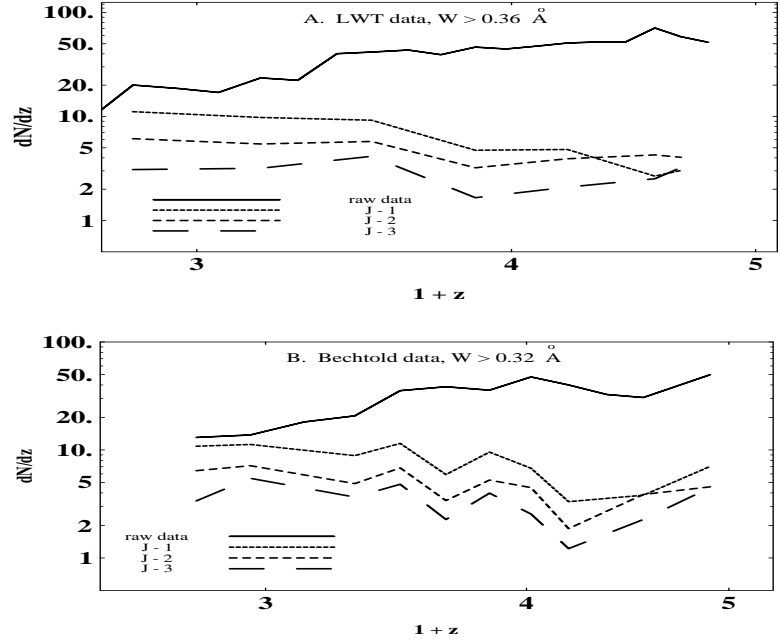,height=4in,width=5.25in}
   \label{fig18}
\vspace{-15mm}
\caption{$dN/dz$ vs $(1+z)$ of A. LWT data of $W>0.36\AA$, B.
JB data of $W>0.32 \AA$. The top curves are given by original
Ly$\alpha$ lines. The lower curves are from the identified clusters on
scales $j=9$, 8 and 7, respectively.}
\end{figure}

We did the same analysis as above with the BGF samples. The results
are shown in Figure 19.  Figure 19a is for BGF
($W > 0.16 \AA$), and 19b for JB ($W>0.16 \AA$). We note first that
the evolution of the number densities of $j=9$, 8 and 7 clusters have
the same trend as real data: $dN/dz$ is decreasing with redshift.
If we fit the curves $dN/dz$ with the power law eq.(16), both the BGF
and JB give about the same index $\gamma$. This shows that the linear
approximation is correct to model the evolutionary trend.
\begin{figure}[htb]
\vspace{5mm}
\epsfig{file=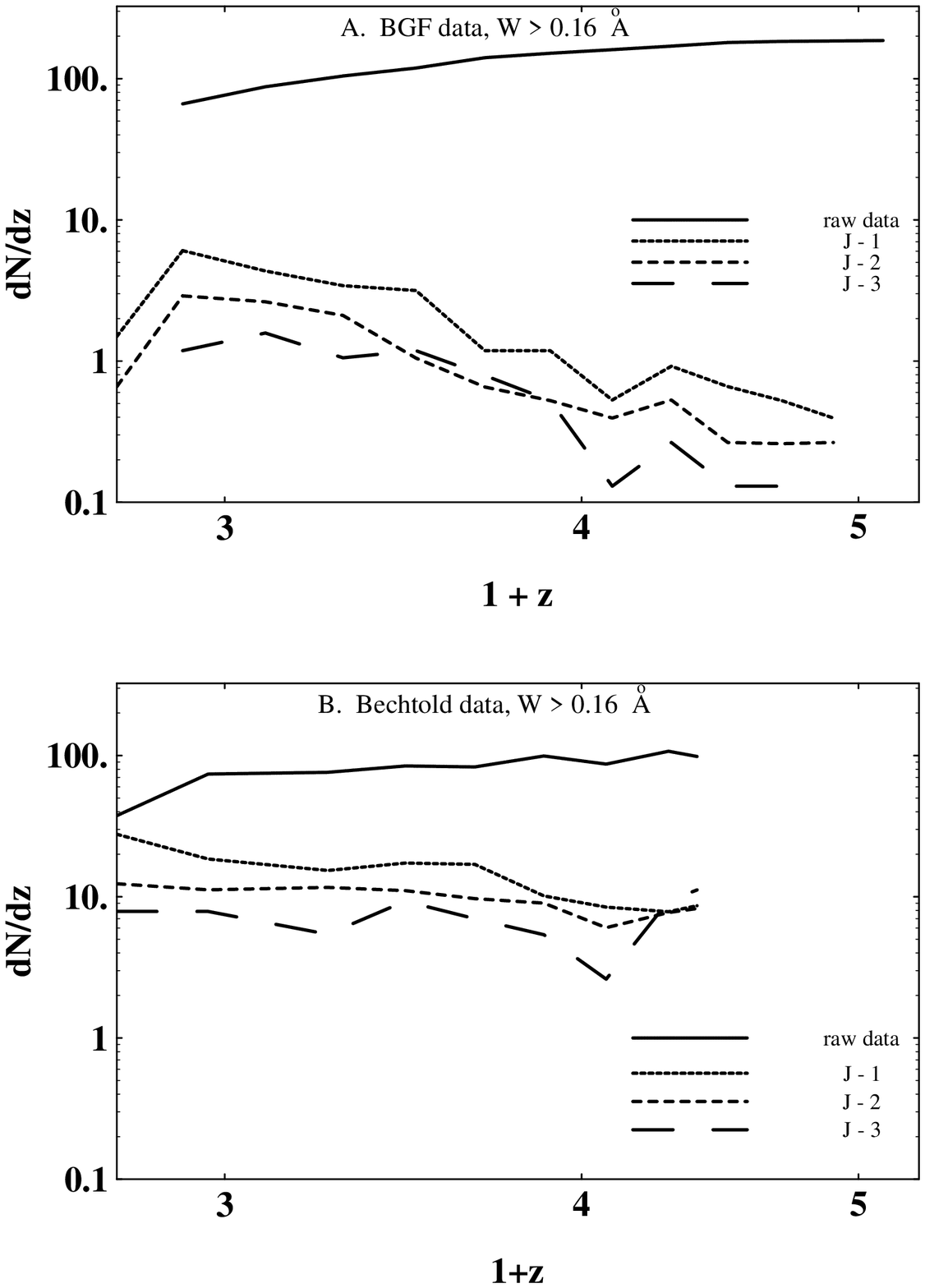,height=4in,width=5.25in}
   \label{fig19}
\vspace{-15mm}
\caption{$dN/dz$ vs $(1+z)$ of A. the BGF sample of LCDM with $W>0.16\AA$,
B. JB data of $W>0.16 \AA$. Both top curves are given by original
Ly$\alpha$ lines. The lower curves are from the identified clusters
on scales $j=9$, 8 and 7. The amplitudes of the BGF curves of $j=9$,
8 and 7 are much lower than the corresponding amplitudes
of the Bechtold data.}
\end{figure}

Yet, the values of $dN/dz$ for the $j=9,$ 8 and 7 clusters of the BGF
data are less than JB's results by a factor of about 5. Considering that
both LWT and JB have about the same number density of Ly$\alpha$ lines (top
curves of Figures 19a and b), the different number density of their clusters
should not be fully given by the uncertainty of current observations.
The discrepancy is due to the fact that the linear simulation
underestimated clustering on large scales.

\setcounter{enumi}{7}
\setcounter{equation}{0}

\section{Miscellaneous topics}

Topics included in this section should not be considered less
important, but rather, less developed.

\smallskip

\noindent{\it 1. Scale-dependence of bias}

\smallskip

Bias is introduced to describe how the spatial distribution of objects
like galaxies and galaxy clusters is related to that of the underlying mass.
Theoretically, linear bias $b$ for a given type of object is defined as
\begin{equation}
\epsilon(x)_{object} = b \epsilon(x)_{mass}.
\end{equation}
Applying the DWT eq.(7.1) gives
\begin{equation}
(\epsilon_{j,l})_{objects} = b (\epsilon_{j,l})_{mass}.
\end{equation}
Considering the density field is homogeneous, both $(\epsilon_{j,l})_{objects}$
and $(\epsilon_{j,l})_{mass}$ should be independent of index $l$.
In eq.(7.2), one can replace $\epsilon_{j,l}$ by its average
$\sum_{l=0}^{2^j-1} |\epsilon_{j,l}|/2^j$. Generally,
$(\epsilon_{j,l})_{objects}$ and $(\epsilon_{j,l})_{mass}$ are
$j$-dependent, and therefore their ratio is also scale-dependent. This is true
even for a linear bias, i.e.,  the factor $b$ is generally scale-dependent.
As a consequence,
it would be better to define the bias factor of the distribution of object
I with respect to object II by FFCs as follows
\begin{equation}
b_j= \frac{(\sum_{l}|\tilde\epsilon_{j,l}|)_{I}}
{(\sum_{l}|\tilde\epsilon_{j,l}|)_{II}}
\end{equation}
where the subscripts $I$ and $II$ denote the FFCs for objects I and II,
respectively.

For instance, figures 17 and 20 show that
$[N_{9}(>2)/N_{8}(>2)]_{0.32} > [N_{9}(>2)/N_{8}(>2)]_{0.16}$, but
$[N_{8}(>2)/N_{7}(>2)]_{0.32} < [N_{8}(>2)/N_{7}(>2)]_{0.16}$,
where the subscripts $0.16$ and $0.32$ denote the clouds of
$W>0.16 \AA$ and $W>0.32 \AA$ of JB samples. This result is not consistent
with $b_{8}=b_{7}$, and therefore, the bias factors $b_j$ of $W>0.16 \AA$
clouds with respect to $W>0.32\AA$ are probably $j$-dependent. Of
course, the uncertainty of the current Ly$\alpha$ forest data is
still large. The $j$-dependence of bias is only a very preliminary
result. However, it already shows that the features of a bias factor, like
scale-dependence, can properly be described by the FFCs.
\begin{figure}[hbt]
\vspace{2cm}
\centering\epsfig{file=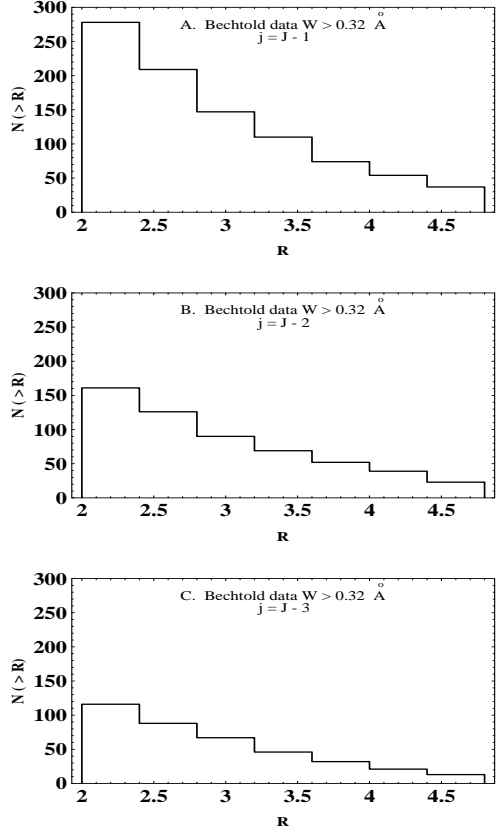, height=4in,width=4in}
   \label{fig20}
\caption{Number of clusters $N_j(>R)$ identified
from the JB $W>0.32\AA$, where $R$ is the richness of the clusters
in units of $\sigma$. A. B. and C. are for scales $j=9$, 8 and 7,
respectively.}
\end{figure}

Bias essentially is one of the environment-dependent effects. On scales equal
to or larger than galaxy clusters, the key parameter of the environment should
be the local background density. The DWT provides a uniform procedure to
measure
the local background density on various scales. It can also transform a
point-like distribution to MFCs on various scales, and reconstruct
the density field on various scales. Environmental effects can
be revealed by the correlation with MFCs. The cross-correlation between
MFC and FFC is particularly important, since it describes how clustering
depends on environment.

\smallskip

\noindent{\it 2. Extraction of fractal}

\smallskip

Observational data indicates that the clustering of galaxies seems to be
scale-free. On the other hand, a pure fractal distribution contradicts with
the observed angular correlations on large scales. A complete picture of the
LSS may need to incorporate fractal structures into a homogeneous
background (Luo and Schramm 1994). However, on which scale will the
fractal features end, and turn to a homogeneous distribution? This issue
is a subject in debate. Statistically, we need a method of
extracting fractal structures from a homogeneous random background, and
detecting the scale of the fractal ending. Since the bases for the DWT
are self similar, the choice of the DWT for fractal analysis is a natural
one.

To illustrate this point, we consider a Gaussian field  with small
self-similar (multifractal) structures added.
In a conventional multifractal analysis a distribution is first expressed
as, for example, eq.(2.10). The scaling behavior is detected by
\begin{equation}
\sum_{l}|\epsilon_{j,l}|^q \sim 2^{j\tau(q)}.
\end{equation}
For a Gaussian background, $\tau_{Gaussian}(q)=q-1$.  Since the amplitudes
$\epsilon_{j,l}$ reflect absolute values, this scaling approach is quite
insensitive to the small self-similar component when the large Gaussian
background dominates. On the other hand, the FFCs $\tilde\epsilon_{j,l}$
focus on
differences between neighboring parts of the distribution, and therefore,
the smooth Gaussian background drops out for small scales $l$ and the
FFCs are only determined by the selfsimilar component. Hence,
the fractal is easily  extracted by
\begin{equation}
\sum_{l}|\tilde\epsilon_{j,l}|^q \sim 2^{-j\beta(q)}.
\end{equation}
The fractal component can be detected if $\beta(q)$ is different
from $(2q-1)$. For large $j$ Gaussian behavior take over again.
One may then be able to find the scale at which the distribution
 transforms from a fractal to a Gaussian distribution.

\smallskip

\noindent{\it 3. Hierarchical clustering and scale-scale correlations}

\smallskip

Besides statistical descriptions, the DWT representation could be valuable
for dynamical study. Dynamics is certainly representation-independent.
Dynamical solutions found in Fourier representation can be
found by a DWT mode expansion. Practically, however, different representation
are
not equivalent because we cannot calculate the mode-mode coupling (or
correlations) on all orders, but only on a few lower orders. These lower
orders are
different for different mode decompositions. In other words, different
bases will reveal different aspects of the LSS dynamics. Different
selection of the bases functions is necessary to study different aspects
of the LSS. To illustrate this point, let's study so-called merging process
in structure formation.

In the standard scenario of structure formation, larger dark matter halos are
generally considered to form hierarchically from the clustering of smaller
halos.
One can trace the histories of dark matter halos on the assumption
that there is a relation between the halos on different scales. In order to
exploit the DWT's ability to extract information of the merging process,
several toy models of textures have been investigated. These include the
p-model, p-model with random branching, $\alpha$-model, QCD-motivated
cascade model, etc. For instance, similar to the Block model of merge trees
(Cole, 1991), the p-model assumes that mass $M$ in a $L$-th order
(scale) object is  split unequally into two $L/2$-th objects.
The process repeats, cascading down through scales of length $L/4$,
$L/8$,.. till the scale $L/n$. The final distribution generated by p-model
is a number density field of $L/n$-th objects. The two-point correlation
function of the $L/n$-th objects shows the same power-law behavior as usually
found in galaxy and other LSS samples.

However, the two-point correlation based on monoscale expansion does not
take the hierarchical process into account. In fact, the power-law behavior
of the two-point correlation can be realized in many models.
It is not an optimal choice to detect the merging trees. A better choice is
the correlation between FFCs on different
scales. One can show that for the p-model, the scale-scale correlations
of FFCs is completely
compressed into the diagonal (Greiner, Lipa, \& Carruthers 1995). That is,
the scale-scale correlation
between FFCs can provide information on how larger structures with
substructures living inside are formed. In this sense, the higher order
FFC correlation are sensitive to the merging dynamics of the LSS.

\section{Outlook}

The wavelet transform is a very new technique in physics.
As with many other mathematical methods being introduced into physics, the
problem in the first phase of development is a lack of feeling for the
physical meaning of the relevant mathematical quantities. In the mind
of physicists, the Fourier coefficient is not only a result of a
mathematical transform, but directly ``seen" as an excited mode, a
``particle" with a given momentum, the energy of the model etc. On the other
hand, the MFCs and FFCs are far from becoming accustomed notions.

This being the case, we should first try to search for a better understanding
of the physical meaning of the wavelet coefficients. Fortunately, the wavelet
community in different fields of physics has shared much with each other.
The DWT is being rapidly introduced in physics including turbulence
(Farge 1992), multiparticle dynamics (Greiner et al. 1996, Huang et al.
1996), disordered solid state systems (Kantelhardt, Greiner \& Roman 1995)
and quantum algebras (Ludu \& Greiner 1995). In the context of LSS, the wavelet
transform certainly opens new possibilities. The results reviewed here
already demonstrate some of the superior properties of the wavelet transform
in revealing the physics of LSS.  As we become more familiar with the technique
and start to build intuition about the MFCs and FFCs, the wavelet transform
should reveal yet more information about LSS.

\newpage

\appendix

\setcounter{equation}{0}

\renewcommand{\theequation}{A\arabic{equation}}

\begin{center}
\begin{Large}

{\bf Appendix}

\end{Large}
\end{center}

\section{Relationship between Fourier coefficients and FFCs}

 From eq.(2.41), we have
\begin{eqnarray*}
\tilde{\epsilon}_{j,l+K} & = &
   \int_{-\infty}^{\infty} \epsilon(x) \psi_{j,l+K}(x)dx \\
   & = &
   \int_{-\infty}^{\infty} \epsilon(x)
   \left( \frac{2^j}{L} \right )^{1/2} \psi(2^jx/L-l-K) dx
\end{eqnarray*}
\begin{equation}
\ \ \ \ \ \ \ \ \ \  =  \
   \int_{-\infty}^{\infty} \epsilon(x' + 2^{-j}KL)
   \left( \frac{2^j}{L} \right )^{1/2} \psi(2^jx'/L-l) dx'
\end{equation}
here we make a change of variable $x'=x - 2^{-j}KL$. Therefore,
when $2^{-j}K$ is an integer, i.e. $K=2^jm$, eq.(A1) is
\begin{equation}
\tilde{\epsilon}_{j,l+2^ jm} =
  \int_{-\infty}^{\infty} \epsilon(x' + mL)
   \left( \frac{2^j}{L} \right )^{1/2} \psi(2^jx'/L-l) dx' =
  \tilde{\epsilon}_{j,l}
\end{equation}
which shows that the FFC, $\tilde{\epsilon}_{j,l}$, are periodic in $l$ with
period $2^j$.

Substituting the wavelet expansion of $\epsilon(x)$, i.e. eq.(2.49) into
eq.(2.51), we have
\begin{equation}
\epsilon_n= \frac{1}{L}\int_0^{L}
 \left [ \sum_{j=0}^{\infty} \sum_{l= - \infty}^{\infty}
  \tilde{\epsilon}_{j,l} \psi_{j,l}(x) \right ] e^{-i2\pi nx/L}dx
\end{equation}
Using eq.(A2), eq.(A3) becomes
\begin{eqnarray*}
\epsilon_n & = & \frac{1}{L}\int_0^{L}
 \left [ \sum_{j=0}^{\infty} \sum_{l= 0}^{2^j-1}
 \sum_{m=-\infty}^{\infty}
  \tilde{\epsilon}_{j,l} \psi_{j,l+2^jm}(x) \right ] e^{-i2\pi nx/L}dx \\
  & = & \frac{1}{L}
  \sum_{j=0}^{\infty} \sum_{l= 0}^{2^j-1} \tilde{\epsilon}_{j,l}
 \sum_{m=-\infty}^{\infty}
  \int_0^{L} \left(\frac{2^j}{L}\right)^{1/2}
  \psi(2^jx/L - l -2^jm)  e^{-i2\pi nx/L}dx \\
  & = & \frac{1}{L}
  \sum_{j=0}^{\infty} \sum_{l= 0}^{2^j-1} \tilde{\epsilon}_{j,l}
  \int_{-\infty}^{\infty} \left(\frac{2^j}{L}\right)^{1/2}
  \psi(2^jx'/L - l ) e^{-i2\pi nx'/L}dx' \\
   & = & \frac{1}{L}
  \sum_{j=0}^{\infty} \sum_{l= 0}^{2^j-1} \tilde{\epsilon}_{j,l}
  \int_{-\infty}^{\infty} \psi_{j,l}(x') e^{-i2\pi nx'/L}dx'
\end{eqnarray*}
\begin{equation}
\ \ \ \ \  = \ \  \frac{1}{L}
  \sum_{j=0}^{\infty} \sum_{l= 0}^{2^j-1} \tilde{\epsilon}_{j,l}
  \hat{\psi}_{j,l}(n)
\end{equation}
This is eq.(2.58). An alternative form, which uses the Fourier transform
of the basic function $\psi(x)$ rather than $\psi_{j,l}(x)$, can be
derived from eq.(A4) as follows
\begin{eqnarray*}
\epsilon_n  & = & \frac{1}{L}
  \sum_{j=0}^{\infty} \sum_{l= 0}^{2^j-1} \tilde{\epsilon}_{j,l}
  \int_{-\infty}^{\infty} \psi_{j,l}(x) e^{-i2\pi nx/L}dx \\
   & = & \frac{1}{L}
  \sum_{j=0}^{\infty} \sum_{l= 0}^{2^j-1} \tilde{\epsilon}_{j,l}
  \int_{-\infty}^{\infty} \left ( \frac{2^j}{L} \right )^{1/2}
  \psi(2^jx/L-l ) e^{-i2\pi nx/L}dx \\
  & = & \frac{1}{L}
  \sum_{j=0}^{\infty} \sum_{l= 0}^{2^j-1}
  \left ( \frac{2^j}{L} \right )^{-1/2}
  \tilde{\epsilon}_{j,l} e^{-i2 \pi nl/2^j}
  \int_{-\infty}^{\infty} \psi(\eta) e^{-i2\pi n\eta/2^j}d\eta
\end{eqnarray*}
\begin{equation}
\ \ \ \ \ \ = \
  \sum_{j=0}^{\infty} \sum_{l= 0}^{2^j-1}
  \left ( \frac{1}{2^jL} \right )^{1/2}
  \tilde{\epsilon}_{j,l} e^{-i2 \pi nl/2^j} \hat{\psi}(n/2^j)
\end{equation}
This is eq.(2.59).


\section{Parseval theorem of the DWT}

\setcounter{equation}{0}

\renewcommand{\theequation}{B\arabic{equation}}


 From the expansion (2.49) we have
\begin{equation}
\int_0^L |\epsilon(x)|^2 dx  =
  \sum_{j, j'= 0}^{\infty} \sum_{l, l'= - \infty}^{\infty}
  \tilde{\epsilon}_{j,l}   \tilde{\epsilon}_{j',l'}
  \int_0^L \psi_{j,l}(x) \psi_{j',l'}(x) dx
\end{equation}
Considering the periodicity (A2), eq.(B1) can be rewritten as
\begin{eqnarray*}
\lefteqn{\int_0^L |\epsilon(x)|^2 dx  } \\
  &  & =
  \sum_{j,j'= 0}^{\infty} \sum_{l = 0}^{2^j-1} \sum_{m=-\infty}^{\infty}
  \sum_{l' = 0}^{2^{j'}-1} \sum_{m'=-\infty}^{\infty}
  \tilde{\epsilon}_{j,l}   \tilde{\epsilon}_{j',l'}
  \int_0^L \psi_{j,l+2^jm}(x) \psi_{j',l'+2^{j'}m'}(x) dx \\
  & &  =
  \sum_{j, j'= 0}^{\infty} \sum_{l = 0}^{2^j-1} \sum_{l' = 0}^{2^{j'}-1}
  \tilde{\epsilon}_{j,l}   \tilde{\epsilon}_{j',l'}  \times \\
  & & \ \ \sum_{m=-\infty}^{\infty} \sum_{m'=-\infty}^{\infty}
  \frac{2^j}{L}
  \int_0^L \psi(2^jx/L - l - 2^jm) \psi(2^{j'}x/L -l' - 2^{j'}m') dx \\
 &  & =
  \sum_{j, j'= 0}^{\infty} \sum_{l = 0}^{2^j-1} \sum_{l' = 0}^{2^{j'}-1}
  \tilde{\epsilon}_{j,l}   \tilde{\epsilon}_{j',l'}
  \sum_{m" \equiv (m-m') =-\infty}^{\infty} \sum_{m'=-\infty}^{\infty} \\
 & & \ \
  \frac{2^j}{L}
  \int_0^L \psi(2^jx/L - l - 2^j(m"+m')) \psi(2^{j'}x/L -l' - 2^{j'}m') dx \\
  &  & =
  \sum_{j, j'= 0}^{\infty} \sum_{l = 0}^{2^j-1} \sum_{l' = 0}^{2^{j'}-1}
  \tilde{\epsilon}_{j,l}   \tilde{\epsilon}_{j',l'}  \times \\
  & & \ \ \sum_{m" =-\infty}^{\infty} \sum_{m'=-\infty}^{\infty}
  \frac{2^j}{L}
  \int_0^L \psi(2^j(x/L-m') - l - 2^jm") \psi(2^{j'}(x/L-m') -l') dx \\
  &  & =
  \sum_{j, j'= 0}^{\infty} \sum_{l = 0}^{2^j-1} \sum_{l' = 0}^{2^{j'}-1}
  \tilde{\epsilon}_{j,l}   \tilde{\epsilon}_{j',l'}  \times \\
  &  &  \ \
  \sum_{m" = -\infty}^{\infty}  \frac{2^j}{L}
  \int_{-\infty}^{\infty} \psi(2^jx/L - l - 2^jm") \psi(2^{j'}x/L-l') dx \\
  &  & =
  \sum_{j, j'= 0}^{\infty} \sum_{l = 0}^{2^j-1} \sum_{l' = 0}^{2^{j'}-1}
  \tilde{\epsilon}_{j,l}   \tilde{\epsilon}_{j',l'}
  \sum_{m" = -\infty}^{\infty}
  \int_{-\infty}^{\infty} \psi_{j,l+2^jm"}(x) \psi_{j', l'}(x) dx
\end{eqnarray*}
\begin{equation}
\ \ \ \ \  = \
  \sum_{j, j'= 0}^{\infty} \sum_{l = 0}^{2^j-1} \sum_{l' = 0}^{2^{j'}-1}
  \tilde{\epsilon}_{j,l}   \tilde{\epsilon}_{j',l'}
  \sum_{m" = -\infty}^{\infty} \delta_{j,j'} \delta_{l+2^jm",l'}
\end{equation}
 From $\delta_{j,j'}$, $j'$ should be equal to $j$, and then
$l'<2^j$. Therefore, $\delta_{l+2^jm",l'}$ requires $m"=0$ and $l=l'$.
We have then the Parseval theorem (4.3).

\newpage

\section{References}

\bigskip

\ref Adler, R.J. 1981, {\it The Geometry of Random Field}, (New York, Wiley)
\ref Barbero, J.F., Dominguez, A., Goldman, T. \& P\'erez-mercader, J.
     1996, Los Alamos preprint LAEFF-96/15
\ref Berera, A.\& Fang, L.Z. 1994, Phys. Rev. Lett, 72, 458
\ref Bechtold, J. 1994, Astrophys. J. Supp. 91, 1 (JB)
\ref Bechtold, J., Crotts, P.S., Duncan, R.C. \& Fang Y. 1994,
     Astrophys. J. 437, L83
\ref Bi, H.G., Ge, J. \& Fang, L.Z. 1995, Astrophys. J. 452. 90, (BGF)
\ref Carruthers, P. 1995, XXIII International Symposium on Multiparticle
Dynamics, Aspen Colorado, World Scientific, Singapore, to be published.
\ref Chui, C.K. 1992, {\it Wavelets: A Tutorial in Theory and Applications},
Academic Press
\ref Cole, S. 1991, ApJ. 367, 45.
\ref Daubechies, I. 1988, Comm. Pure. Appl. Math. 41, 909.
\ref Daubechies, I. 1992, {\it Ten Lectures on Wavelets}, (SIAM)
\ref Dinshaw, N., Foltz, C.B., Impey, Weymann, R. \& Morris, S.L.
     1995, Nature, 373, 223
\ref Efstathiou, G., Kaiser, N., Saunders, W., Lawrence, A.,
    Rowan-Robinson, M., Ellis, R.S. \& Frenk, C.S. 1990,
    Mon. Not. R. Astr. Soc. 247, 10p
\ref Escalera, E. \& Mazure, A. 1992, Astrophys. J. 388, 23
\ref Escalera, E., Slezak, E. \& Mazure, A. 1992, Ast.
    \& Astrophys. 264, 379
\ref Fan, Z.H. \& Bardeen, J.M. 1995, Phys. Rev.  D51, 6714
\ref Fang, L.Z. 1991, Astr. \& Astrophys. 244, 1.
\ref Fang, Y.H., Duncan, R.C., Crotts, A.P.S. \& Bechtold, J.,
     Astrophys. J., 462, 77
\ref Farge, M. 1992, Ann. Rev. Fluid Mech., 24, 395
\ref Greiner, M., Lipa, P. \& Carruthers, P. 1995, Phys. Rev. E51, 1948
\ref Greiner, M., Giesemann, J., Lipa, P. \& Carruthers, P. 1996,
     Z. Phys. C.69, 305
\ref Huang, Z. Sarcevic, I., Thews, R. \& Wang, X.N. 1996, Phys. Rev. D.
     54, 750
\ref Ivanov, A.V. \& Leonenko, N.N. 1989, {\it Statistical analysis of Random
Field},  Klumer Academic Pub.
\ref Kaiser, G. 1994, Applied \& Computational Harmonic Analysis, 1, 246.
\ref Kaiser, N. \& Peacock, J.A. 1991, Astrophys. J. 379, 482
\ref Kantelhardt, J., Greiner, M. \& Roman, E. 1995, Physica, A220, 219
\ref Kolb, E.W. \& Turner, M.S. 1989, {\it The Early Universe},
    Addison-Wesley Pub. Co.
\ref Liu, X.D. \& Jones, B.J.T. 1990, Mon. Not R. Astr. Soc. 242, 678
\ref Lu, L., Wolfe, A.M., \& Turnshek, D.A. 1991, Astrophys. J. 367, 19 (LWT)
\ref Ludu, A. \& Greiner, M. 1995, ICTP-Internal Report IC/95/214
\ref Luo, X. \& Schramm, D. 1992, Science, 256, 513
\ref Mallat, S. 1989, Trans. Am. Math. Soc. 315, 69
\ref Mallat, S. \& Zhong, S. 1990, Courant Inst. Tech. Rep. No. 483,
\ref Martinez, V.J. Paredes, S. \& Saar, E. 1993,
  Mon. Not. R. Astr. Soc. 260, 365
\ref Meyer, Y. 1988, Congr. Int. Phys. Math. July
\ref Meyer, Y. 1992, {\it Wavelets},  (SIAM)
\ref Meyer, Y. 1993, {\it Wavelets: Algorithms and Applications}, SIAM
\ref Pando, J, \& Fang, L.Z., 1995, astro-ph/9509032
\ref Pando, J, \& Fang, L.Z., 1996a, Astrophys. J., 459, 1.
\ref Pando, J, \& Fang, L.Z., 1996b, astro-ph/9606005
\ref Peebles, P.J.E. 1980, {\it The Large Scale Structure of the Universe},
    Princeton Univ. Press.
\ref Perivolaropoulos, L. 1994,  Mon. Not. R. Astr. Soc. 267, 529
\ref Press, W.H., Teukolsky, S.A., Vetterling, W.T. \& Flannery, B.P.
    1992, {\it Numerical Recipes}, Cambridge
\ref Slezak, E., Bijaoui, A. \& Mars, G. 1990, Astro \& Astrophys, 227, 301
\ref Vanmarcke, E,  1983, {\it Random Field}, MIT Press.
\ref Yamada, M. \& Ohkitani, K. 1991, Prog. Theor. Phys., 86, 799
\ref Weymann, R. J. 1993, in The Environment and Evolution of
Galaxies, ed. Shull, J. M. \& Thronson, H. A. Jr. 213
\ref Wickerhauser, M.V. 1994, {\it Adapted Wavelet Analysis From Theory to
Software}, A.K. Peters, Mass

\end{document}